\documentclass{emulateapj}
\usepackage{color}

\usepackage{natbib}

\newcommand {\ia}{\'\i }

\begin{document}



%
%
%

%
%
\title{The soft X-ray and narrow-line emission of Mrk\,573 on kiloparcec scales}
\author{O. Gonzalez-Martin\altaffilmark{1,2}, J.A.
Acosta-Pulido\altaffilmark{3,4}, A.M. Perez Garcia\altaffilmark{3,4}, 
C. Ramos Almeida\altaffilmark{5}}
\altaffiltext{1}{IESL, Foundation for Research and Technology, 71110, Heraklion, Crete, Greece.             
 omaira@physics.uoc.gr}
\altaffiltext{2}{University of Crete, Department of Physics, Voutes, 71003 Heraklion, Crete, Greece.}

\altaffiltext{3}{Instituto de Astrof\'{i}sica de Canarias (IAC), 
              C/V\'{i}a L\'{a}ctea, s/n, E-38205, La Laguna, Tenerife, Spain.
jap@iac.es, apg@iac.es}
\altaffiltext{4}{Departamento de Astrof{\'i}sica, Universidad de La Laguna, E-38205 La Laguna, Tenerife, Spain.}
\altaffiltext{5}{Department of Physics \& Astronomy, University of Sheffield, Sheffield, S3 7RH, UK, C.Ramos@sheffield.ac.uk}

\begin{abstract}

We present a study of the circumnuclear region of the nearby Seyfert galaxy Mrk~573 using 
\emph{Chandra}, \emph{XMM-Newton} and \emph{Hubble Space Telescope} (\emph{HST}) data. 
We have studied the morphology of the soft ($\sf{< 2~keV}$) X-rays comparing it with the [O III] 
and H$\rm{\alpha}$ \emph{HST} images.
The soft X-ray emission is resolved into a
complex extended region. The X-ray morphology shows
a biconical region extending up to 12 arcsecs (4 kpc) in projection from
the nucleus. A strong correlation between the X-rays and the highly ionized gas
seen in the [O III]$\mathrm{\lambda 5007 }$~\AA{} image is reported. 
Moreover, we have studied the line intensities detected with the \emph{XMM--Newton} Reflection Grating Spectrometer 
(RGS) and used them to fit the low resolution EPIC/\emph{XMM--Newton} and ACIS/\emph{Chandra} spectra.
The RGS/\emph{XMM--Newton} spectrum is dominated 
by emission lines of C VI, O VII, O VIII, Fe XVII, and Ne IX, among others highly ionized species.
A good fit is obtained using these emission lines found in the RGS/\emph{XMM--Newton} spectrum as 
a template for \emph{Chandra} spectra of the nucleus and extended emission, coincident with 
the cone-like structures seen in the [O III]/H$\rm{\alpha}$ map. 
The photoionization model Cloudy provides a reasonable fit for both the nuclear region and the 
cone-like structures showing that the dominant excitation mechanism is photoionization. 
For the nucleus the emission is modelled using two phases: a high ionization [log(U)=1.23] and a 
low ionization [log(U)=0.13]. For the high ionization phase the transmitted and reflected component are in
a ratio 1:2, whereas for the low ionization the reflected component dominates. 
For the extended emission, we successfully reproduced the emission with two phases. 
The first phase shows a higher ionization parameter for the NW (log(U)=0.9) 
than for the SE cone (log(U)=0.3). Moreover, this phase is transmission dominated for the SE cone
and reflection dominated for the NW cone. The second phase shows a low ionization parameter (log(U)=-3)
and is rather uniform for NW and SEcones and equally distributed in reflection and transmission components.
In addition, we have also derived the optical/infrared spectral energy distribution (SED) of the nucleus 
from high spatial resolution images of Mrk~573.
The nuclear optical/infrared SED of the nucleus has been modeled by a clumpy torus model. The torus
bolometric luminosity agrees very well with the AGN luminosity inferred from the observed hard X-ray spectrum. 
The optical depth along the line of sight expected from the torus modeling indicates 
a high neutral hydrogen column density in agreement with the classification 
of the nucleus of Mrk\,573 as a \emph{Compton-thick} AGN. 

\end{abstract}

   \keywords{galaxies:active - galaxies:nuclei - galaxies:Seyfert - galaxies: individual (Mrk~573) - 
infrared:galaxies}

\section{Introduction}\label{sec:intro}

Visible signatures of the direct interaction between the active galactic nuclei (AGN) and their host galaxies
include kpc-scale [O III]- and H$\sf{\alpha}$-emitting regions, the so-called extended narrow-line region (ENLR), which is
observed in many nearby Seyfert galaxies \citep{Schmitt03,Veilleux03,Whittle04,Ramos06}. 
Understanding the AGN-host galaxy interaction and feedback is 
crucial for the study of both galaxy and AGN evolution \citep{Silk98,Kauffmann03,Hopkins06,Schawinski07,Schawinski09}. 

One of the most promising ways to study these interactions is through soft X-rays 
\citep{Crenshaw99,Crenshaw03,Dai08}.  
In the unified picture of AGN \citep{Antonucci93} 
the soft X-ray spectra of Type-2 Seyfert galaxies are expected to be affected by emission and scattering from
the medium, which is also strongly influenced by the nuclear continuum. However, soft X-rays 
can also be produced through mechanical heating, as in shocks driven by supernova explosions
in nuclear star-forming regions. Indeed, it is quite plausible that both effects
are important (e.g.\ Mrk\,3; \citealt{Sako00}). This soft X-ray emission provides the
opportunity to obtain an X-ray diagnostic for the physical properties of the interacting interstellar medium [ISM]
\citep{Young01,Yang01,Ogle00,Ogle03,Bianchi06,Evans06,Kraemer08}. 

The galaxy Mrk~573 has been extensively studied by many authors. Its active nucleus is 
hosted in an (R)SAB(rs)0+ galaxy.\footnote{The redshift of the 
galaxy has been measured to be  $z=0.017$ \citep{Ruiz05}, which implies a
physical scale of 333 pc~arcsec$^{-1}$, using H$_{0}$ = 75 km s$^{-1}$ Mpc$^{-1}$.} 
Mrk\,573 is well-known for its extended, richly structured circumnuclear 
emission-line regions \citep{Tsvetanov92}. It has long been known to show a biconical structure
and bright arcs and knots of line-emitting gas \citep{Ferruit99,Quillen99}
that are strongly aligned and interacting with a kiloparsec-scale low-power
radio outflow \citep{Pogge93,Falcke98,Ferruit99}. 
\citet{Schlesinger09} have recently studied the STIS/{\it HST} spectrum, finding that
the ENLR optical spectrum is consistent with photoionization by the AGN.

In their study of the ACIS/{\it Chandra} spectrum of Mrk\,573 \citet{Guainazzi05} showed that 
it is consistent with a {\it Compton-thick} source (i.e.\ ${\sf N_H > 1.6 \times 10^{-24} cm^{-2}}$),
showing a large EW of the FeK${\sf \alpha}$ emission line and a steep photon spectral index ($\Gamma$ = 2.7$\pm$0.4). 
These characteristic X-ray properties, together with high-quality LIRIS near-infrared  spectroscopy,  
allowed this galaxy  to be reclassified as an obscured narrow-line Seyfert 1 \citep{Ramos08,Ramos09}.

RGS/{\it XMM--Newton} high resolution observations of Mrk\,573 were reported among a sample of 69 objects 
by \citet{Guainazzi08}, whose spectrum shows strong emission coming 
from the O VII triplet and Ly${\it \alpha}$
O VIII features (see their figure~1), but only few emission line fluxes were reported in their work. 
However, the analysis of the extended soft X-ray emission had not been previously reported in the literature. 


In this paper, we discuss the extended kpc-scale emission of Mrk\,573 using ACIS/{\it Chandra} high 
resolution images, RGS/{\it XMM--Newton} high resolution spectra and {\it HST} imaging. 
While RGS/\emph{XMM--Newton} data 
give the opportunity to use emission line diagnostics to understand 
the nature of this emission, {\it Chandra} high resolution images give the opportunity to 
spatially resolve the emission and study their properties separately. {\it HST} images 
allow us to establish the connection between the optical structure and the soft X-ray
emission. The paper is presented as follows. Section 2 describes the data reduction and 
processing. We present the study of the  circumnuclear morphology and the X-ray spectral analysis in 
Sections 3 and 4, respectively. The origin of this soft X-ray emission is discussed in Section 5 and the 
optical/infrared spectral energy distribution is studied in Section 6. Finally, conclusions and the overall picture for 
Mrk\,573 are presented in Section 7. 

\section{Observations and data processing}\label{sec:obs}

\subsection{XMM-Newton data}\label{sec:obs:xmm}

We retrieved from the HEASARC archive the \emph{XMM--Newton} observation of Mrk\,573
taken on 2004 January 15 (ObsID 0200430701). 
RGS data were processed with the standard RGS pipeline 
processing chains incorporated in the \emph{XMM--Newton} SAS v.8.0.1 \citep{Gabriel04}. Dispersed source 
and background spectra (using blank field event lists) were extracted with automatic RGS extraction tasks. 
The net exposure time is 9 ks after flare removal and the net count rate is 
0.21~$count~s^{-1}$

In this paper we also used lower resolution spectrum from the EPIC pn camera \citep{Struder01}. Source
regions were extracted within a circular region of 25 arcsec\footnote{This radius contains the 80\% (85\%)
of the Point Spread Function (PSF) at 1.5 keV (9.0 keV) for an on-axis source with EPIC pn instrument.} radius centered at 
the position given by NED\footnote{http://nedwww.ipac.caltech.edu/}. We also used {\sc eregionanalise} 
SAS task to compute the best centroid for our source. The difference between the best fit centroid and the
NED position is 2 arcsec (P.A. $\rm{126^o}$). It implies $\rm{\sim}$1\% of error on the final flux, according to 
the PSF of EPIC pn instrument.
Background was selected from a circular region in the same chip as 
for the source region, excluding point sources. Regions were extracted by using the {\sc evselect} task and pn 
redistribution matrix, and effective areas were calculated with the 
{\sc rmfgen} and {\sc arfgen} tasks, respectively.
We also binned the EPIC/pn spectrum to give a minimum of 20 counts per 
bin before background subtraction to be able to use the $\sf{\chi^{2}}$ as the fit 
statistics using the {\sc grppha} task. Note that the background is only 2.4\% of the total number of 
counts ($\rm{\sim50~counts}$). Thus, the spectral binning can be done before background substraction in this source.

\subsection{Chandra data}\label{sec:obs:chandra}

Mrk~573 was observed by \emph{Chandra} on 2006 November 11.
Level 2 event data from the ACIS instrument were extracted from the
\emph{Chandra} archive\footnote{http://cda.harvard.edu/chaser/} (ObsID 7745). 
The data were reduced with the {\sc ciao 3.4}\footnote{http://asc.harvard.edu/ciao} 
data analysis system and the 
\emph{Chandra}  Calibration Database (caldb 3.4.0\footnote{http://cxc.harvard.edu/caldb/}). 
The exposure time was processed to exclude background flares using the {\sc
lc\_clean.sl} task\footnote{http://cxc.harvard.edu/ciao/download/scripts/} 
in source-free sky regions of the same observation. The net exposure time 
after flare removal is 35~ks and the net count rate is 1.5~$s^{-1}$. 
The nucleus has not significantly piled up.

\emph{Chandra} data include information about the photon energies 
and positions that was used to obtain energy-filtered images and 
to carry out sub-pixel resolution spatial analysis. Although the default
pixel size of the \emph{Chandra}/ACIS detector is $\sf{0.492}$ arcsec, smaller 
spatial scales are accessible as the image moves across the detector
pixels during the telescope dither, therefore sampling pixel scales smaller than 
the default pixel of {\it Chandra}/ACIS detector. This allows sub-pixel binning of the images.
Similar techniques were applied  
for the analysis of \emph{Chandra} observations of, for example, the SN1987A remnant \citep{Burrows00}.

In addition to the high-spatial resolution analysis, we applied 
smoothing techniques to detect the low-contrast diffuse emission.  
We applied the adaptive smoothing CIAO tool {\sc csmooth}, based on the algorithm
developed by \citet{Ebeling06}. 
{\sc csmooth} is an adaptive smoothing tool for images containing 
multiscale complex structures and preserves the spatial signatures
and the associated counts, as well as significance estimates.
A minimum and maximum significance S/N level of 3 and 4, 
and a scale maximum of 2 pixels were used. 

We also performed spectral analysis of the nuclear and extended emission 
using CIAO software. Background regions were defined by source-free apertures around the soft X-ray 
emission regions. Response and 
ancillary response files were created using the CIAO {\sc mkacisrmf} and 
{\sc mkwarf} tools. To be able to use the $\sf{\chi^{2}}$ as the fit 
statistics, the spectra were binned to give a minimum of 20 counts per 
bin before background subtraction using the {\sc grppha} task, included 
in FTOOLS.\footnote{http://heasarc.gsfc.nasa.gov/docs/software/ftools}

\begin{table}[!t]
\begin{center}   
\caption{High spatial resolution nuclear fluxes}
\begin{tabular}{cccc}
\hline \hline
$\lambda_{ref} (\micron)$ & $F_\nu$ (mJy) & Instrument/Filter & Ref. \\
\hline 
0.675  & 0.002    & WFPC2/F675W    & a \\
0.814  & 0.003    & WFPC2/F814W    & a \\
1.100  & 0.036     & NICMOS/F110W   & a \\
1.650  & 0.479	    & NICMOS/F160W   & a \\
2.120  & 3.20       & NSFCam/K'      & b \\
3.510  & 18.8       & NSFCam/L       & b \\ 
4.800  & 41.3       & NSFCam/M       & b \\
10.36  & 177        & T-ReCS/N       & c \\
18.30  & 415        & T-ReCS/Qa      & c \\
\hline
\end{tabular}
\tablecomments{Errors in flux densities are $\sim$10\% for HST fluxes, $\sim$20\% for the 3 m NASA IRTF, 
$\sim$15\% for T-ReCS/N, and $\sim$25\% for T-ReCS/Qa.}
\tablerefs{(a) this work ; (b) \citet{AlonsoHerrero03}; (c)
\citet{Ramos09b}. }
\label{tab:psf}
\end{center}
\end{table}

\subsection{Optical and near-IR data}\label{sec:obs:opir}

In  order to perform a detailed comparison between X-ray and optical cone-like 
structures we have retrieved \emph{HST}/WFPC2 narrow-band optical images, centered at 
5343~\AA ([O III]) and 6510~\AA (H$\alpha$). These images were
obtained within the Cycle 4 program  6332 (PI Wilson) and retrieved from the Hubble Legacy Archive. 
Details of the observations can be found in \citet{Falcke98}. The two images were re-centered assuming that the 
peak in both images corresponds to the same locus \citep{Ferruit99}. The same assumption was used to align optical 
and X-rays images. Note that high accuracy is not needed due to the lower \emph{Chandra} resolution.

We have also investigated the nuclear properties of 
Mrk\,573 in the optical and near-IR ranges, based on high spatial resolution 
broad-band images. We have found optical and near-IR broad-band images taken with the 
WFPC2 and NICMOS cameras on board the \emph{HST}.
In particular, we used F675W and F814W WFPC2 images (program 5746; PI  Machetto), and
F110W and F160W NICMOS images (program 7867, PI Pogge). 
All  data were retrieved from the Hubble Legacy Archive.
Post-pipeline images have been cleaned of cosmic rays using the
IRAF task {\it lacos\_im} \citep{vandokkum01}. 
For the analysis we have first separated the nuclear emission 
from the underlying host galaxy emission. We have applied the two-dimensional image decomposition  
{\it GALFIT} program \citep{Peng02} to fit and subtract the unresolved component (PSF).  
Model PSFs were created using the TinyTim 
software package, which includes the optics of \emph{HST} 
plus the specifics of the camera and filter system \citep{krist93}. 
The filters used here are not contaminated
by line emission except for the case of F110W, which contains strong Pa$\beta$ emission.
We have used the nuclear near-IR spectrum of
Mrk\,573 obtained with the LIRIS spectrograph \citep[see][]{Ramos09} to subtract this
emission line contribution. The Pa$\beta$ nuclear emission (including the broad and narrow
components) amounts  $3.17 \times 10^{-14}$ erg cm$^{-2}$ s$^{-1}$. The corrected values of the flux obtained
for the nuclear component in each filter are given in Table \ref{tab:psf}. 

\section{Circumnuclear morphology}\label{sec:morpho}
 
\begin{figure}[!t]
\begin{center}
\includegraphics[width=1.0\columnwidth]{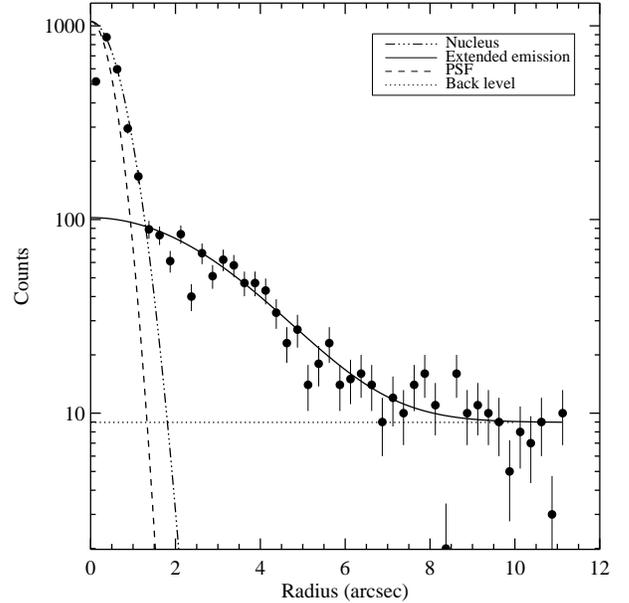}
\caption{\footnotesize{Radial profile of the soft (0.2--2.0 keV) X-ray emission of Mrk\,573 
(black dotted). Continuous lines correspond to the fit to the 
inner ($\rm{< 1.5 arcsecs}$) and outer parts of the radial profile. 
Dashed line shows the radial profile of the simulated 
PSF at $\rm{\sim}$1 keV. The model used to fit the radial profile
is of the form: $y(x)=A\times e^{-x^{2}/2B^{2}}$. An additive constant was added to the 
outer parts of the radial profile fit 
to include the minimum S/N level (dotted line). The FWHM of the inner parts is 0.59 
arcsecs, consistent with the FWHM of the PSF (0.43 arcsecs).}}
\label{fig:PSF}
\end{center}
\end{figure}

\begin{figure*}
\begin{center}
\includegraphics[width=1.0\columnwidth]{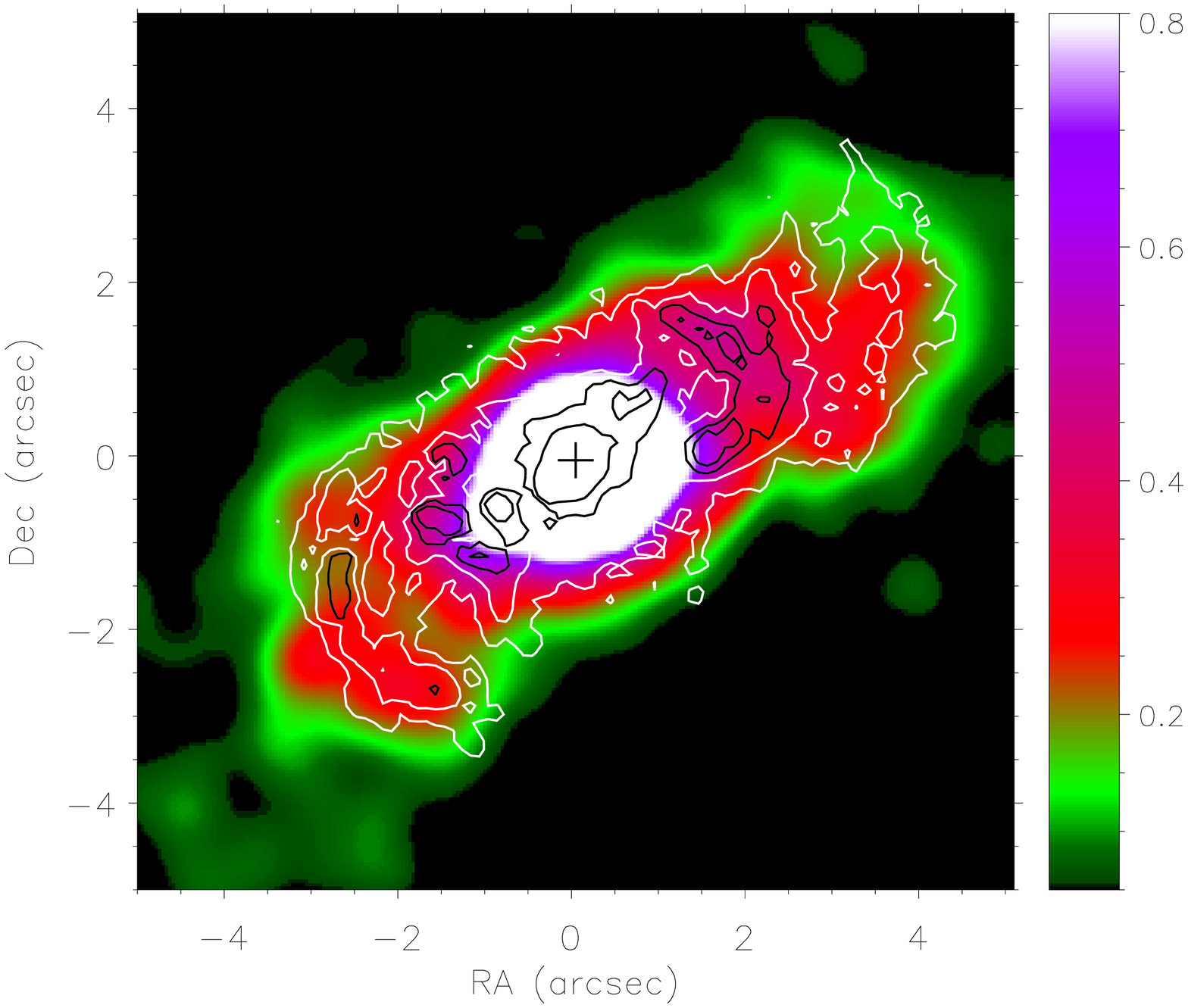}
\includegraphics[width=1.0\columnwidth]{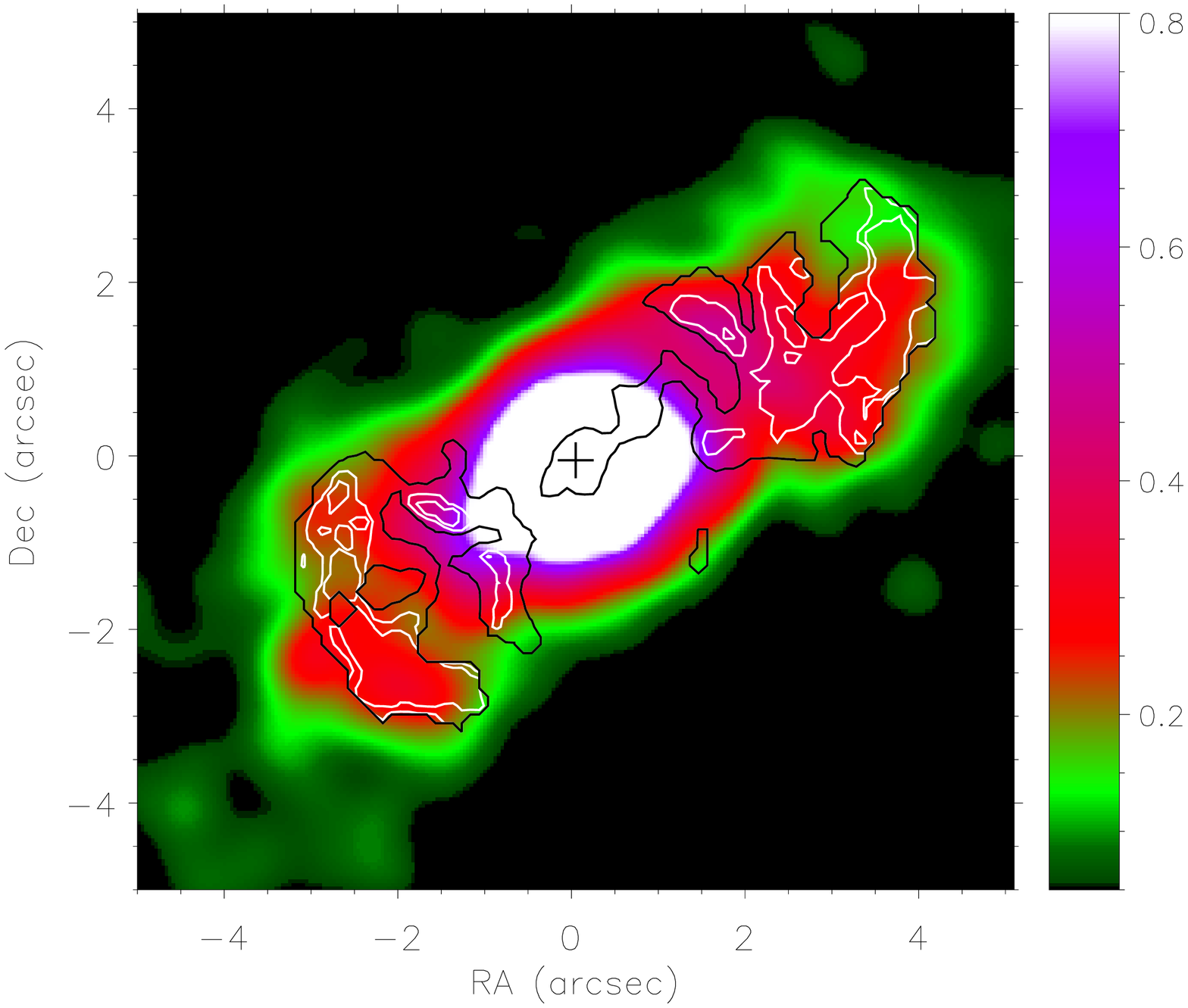}
\caption{\footnotesize{Soft 0.2-2 keV ACIS/{\it Chandra} image of Mrk\,573. {\it
(top)}: [O III] contours overlaid.
{\it (bottom)}: [O III]/$\rm{H\alpha}$ contours overlaid. }
\label{fig:X_OIII}}
\end{center}
\end{figure*}

\begin{figure}[!b]
\includegraphics[width=0.9\columnwidth]{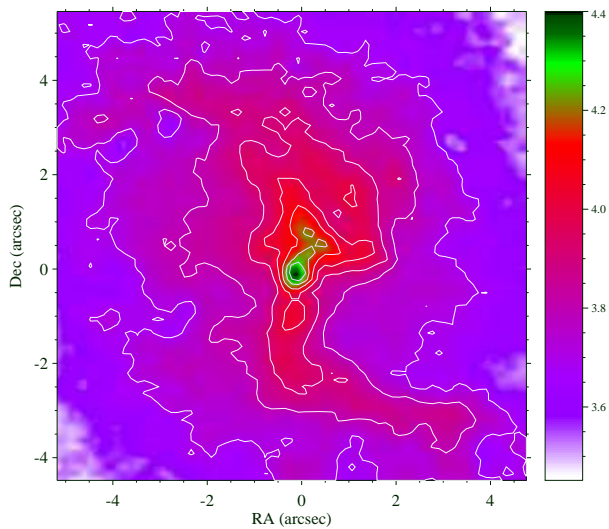}
\caption{\footnotesize{I-H color map of Mrk\,573. Left is the East and up is the North coordinates.}}
\label{fig:color}
\end{figure}

Two broad-band images using {\it Chandra}/ACIS data\footnote{We used {\it Chandra} data for morphological 
analysis since it provides the best X-ray spatial resolution.} were created: (1) soft band at 0.2--2 keV, and (2)
hard band at 2--10 keV. PSF simulations were carried out using information on the spectral
distribution and off-axis location of the system as inputs to ChaRT PSF
simulator.\footnote{http://asc.harvard.edu/chart/index.html} 
Hard band ($\sf{>}$2 keV) shows a point-like morphology, consistent with the FWHM PSF 
simulations, while soft band ($\sf{<}$2 keV) shows an extended morphology (see Figure \ref{fig:PSF}). 
 
The soft X-ray image shows a complex extended emission with a bipolar
structure aligned along PA${\sf \sim 122^{\circ}}$ (Figure \ref{fig:X_OIII}). It extends 8.5 arcsec to the NW and 
6.5 arcsec to the SE. As already mentioned, Mrk\,573 shows a biconical morphology in the [O III] emission,
which resembles that observed in the soft X-rays. 

The ratio of [OIII] to H recombination lines is commonly assumed to be an indicator of the ionization
degree in line emitting photoionized gas.
We have computed the [O III]/$\rm{H\alpha}$ map by dividing the flux-calibrated HST/WFPC2 images in those 
pixels where the signal-to-noise is higher than 3. 
A smoothing algorithm using the median within a box of 0.5 arcsec was then applied. 
Note that this ratio image is only used for morphological comparison and not for any quantitative
analysis since the $\rm{H\alpha}$ image is contaminated with [N II] emission. 
 The use of H$\alpha$ instead of H$\beta$ to compute the ionization map could be affected by reddening 
variations, as shown close to the nucleus of Mrk\,573. However, the extinction map (Fig. \ref{fig:color}) shows 
a very different morphology to that observed in the ratio [O III]/$\rm{H\alpha}$  map. This implies that 
the basic morphology of the optical ionization map cannot be explained by differential extinction, 
despite a certain fraction of the emission line map variations could be due to it.

The X-ray arc-like structure at 10 arcsec to the SE resembles the [O III] emission. However, the 
X-ray emission does not show the bridge between this structure and the inner parts of the SE cone seen in the [O III] 
image (also seen in [O III]/$\rm{H\alpha}$).
The NW structure is coincident with the [O III] emission and strongly 
similar to the [O III]/$\rm{H\alpha}$ morphology (see Figure \ref{fig:X_OIII}). These facts indicate a
link between the soft  X-ray emission and the optical ionized gas, although the detailed structure would depend on the 
small-scale ionization structure of the medium.  

We have also explored the circumnuclear extinction. For this purpose we have used the broad-band  
images after subtracting the unresolved component to build a colour map of the circumnuclear region of Mrk\,573. 
The colour map was created using the F814W and the F160W images, which are equivalent to the $I$ and $H$-bands, respectively. 
These filters were selected because they are not critically contaminated by emission lines, contrary to the colour maps
presented by \citet{Quillen99}.
The $I-H$ map covering the central 10 arcsec is shown in Figure \ref{fig:color}.
The reddest (or darkest in black/white) regions show the location of the strongest obscuration, or equivalently, the highest
dust concentration. 
The colour map reveals a dust lane crossing 
the nucleus in the N--S direction, which bends after 2$''$ resembling incipient
spiral arms. The axis delineated by this structure is 
oriented at about $55^\circ$ with respect to the alignment of the radio structure 
\citep[PA~$\sim 125^\circ$]{Falcke98,Kinney00}. 
The radio and the dust lanes axis should   
be almost perpendicular if the dust lanes were the outer parts of the postulated dusty torus for
Type-2 nuclei. However, this is not necessarily the case, since the dust lane 
is observed at much larger scale than that of the postulated torus, changes in the
dust plane may take place when approaching the inner nucleus.
On the other hand, projection effects could mask the actual orientation of 
both structures, for instance \citet{Tsvetanov92} proposed that the ionization cone 
is very inclined ($35^\circ$) with respect to the plane of the sky. In addition, 
there is also a tongue extending $0\farcs5$ towards the N--NW which ends in two small blobs
that resembles a structure observed in the excitation maps (see Figure \ref{fig:X_OIII}). 
The presence of dust within the ionization cone has been reported before only in NGC~1068 \citep{Bock00}.
Towards the nucleus of the galaxy there is an excess of reddening which can be attributed to
a natural increase in the extinction due to higher dust concentration. 
Assuming an intrinsic colour similar to that observed in the disk of galaxies 
\citep[$\mathrm{I-H} \sim 2$; ][]{Moriondo98}, 
we have estimated a value of ${A_V = 6.5}$ mag, which results in ${N_H = 1.2\times 10^{22}} \mathrm{cm}^{-2}$. 

\begin{table}[!t]
\begin{center}                          
\caption{Emission lines detected in the RGS/{\it XMM-Newton} spectrum.}         
\begin{tabular}{l l l c}   \\     
\hline\hline                 
\multicolumn{1}{c}{Name}  &  \multicolumn{1}{c}{$\lambda_{rest}$}  & \multicolumn{1}{c}{Energy} & 
  \multicolumn{1}{c}{Flux}     \\
 &  \multicolumn{1}{c}{(\AA)}  & \multicolumn{1}{c}{(keV)} &  
\multicolumn{1}{c}{$10^{-5}$}     \\ \hline
CV He$\gamma$        & 33.426 & 0.376 &    10.7$\, _{  0.00}^{  27.3}$  \\
NVI He$\alpha$ (f)   & 29.534 & 0.420 &    0.00$\, _{  0.00}^{  82.7}$  \\
NVI He$\alpha$ (i)   & 29.083 & 0.426 &    2.64$\, _{  0.00}^{  57.6}$  \\
NVI He$\alpha$ (r)   & 28.787 & 0.431 &    0.00$\, _{  0.00}^{  1.85}$  \\
CVI Ly$\beta$        & 28.466 & 0.436 &    3.18$\, _{  0.57}^{  7.10}$  \\
NVII Ly$\alpha$      & 24.781 & 0.500 &    1.03$\, _{  0.00}^{  2.51}$  \\
OVII He$\alpha$ (f)  & 22.101 & 0.561 &    5.17$\, _{  2.33}^{  9.21}$  \\
OVII He$\alpha$ (i)  & 21.803 & 0.569 &    0.00$\, _{  0.00}^{  2.06}$  \\
OVII He$\alpha$ (r)  & 21.602 & 0.574 &    3.76$\, _{  0.96}^{  7.58}$  \\
OVIII Ly$\alpha$     & 18.969 & 0.654 &    2.17$\, _{  1.00}^{  3.70}$  \\
OVII He$\gamma$      & 17.768 & 0.698 &    1.60$\, _{  0.31}^{  3.36}$  \\
FeXVII 3s--2p        & 17.078 & 0.726 &    0.75$\, _{  0.00}^{  1.84}$  \\
  OVII RRC	     & 16.771 & 0.739 &   1.07$\, _{  0.03}^{  2.65}$  \\
FeXVII 3d--2p        & 15.010 & 0.826 &   0.65$\, _{  0.02}^{  1.45}$  \\
 OVIII RRC           & 14.228 & 0.882 &   0.48$\, _{  0.00}^{  1.33}$  \\
NeIX(f)              & 13.698 & 0.905 &   0.22$\, _{  0.00}^{  0.94}$  \\
NeIX(i)              & 13.552 & 0.915 &   0.12$\, _{  0.00}^{  1.10}$  \\
NeIX(r)              & 13.447 & 0.922 &   0.76$\, _{  0.08}^{  1.87}$  \\
\hline
\end{tabular} \\
\tablenotemark{a}{Units are $\rm{ph~cm^{-2}~s^{-1}}$.}													        				 
\label{tab:RGS}      
\end{center}
\end{table}

\section{X-ray spectral analysis}\label{sec:results:spectra}

X-ray spectral analysis of the observed soft X-ray extended emission is crucial to determine the excitation mechanism 
of the plasma, and its relationship to the optical bicone-like structure (see previous section). 
The combination of RGS/{\it XMM-Newton} high spectral resolution and ACIS/{\it Chandra} high spatial 
resolution data is key to achieve this purpose. In this section we describe in detail the methodology 
and main results obtained. 
In Section \ref{sec:origin} we discuss the origin of this extended emission
based on the results presented in this section.
The analysis of the spectral counts was performed using the software package XSPEC 
\citep[version 12.4.0\footnote{http://cxc.heasarc.gsfc.nasa.gov/docs/xanadu/xspec/, }; ][]{Arnaud96}. 

\begin{figure*}
\begin{center}
\includegraphics[height=0.9\textwidth,angle=90]{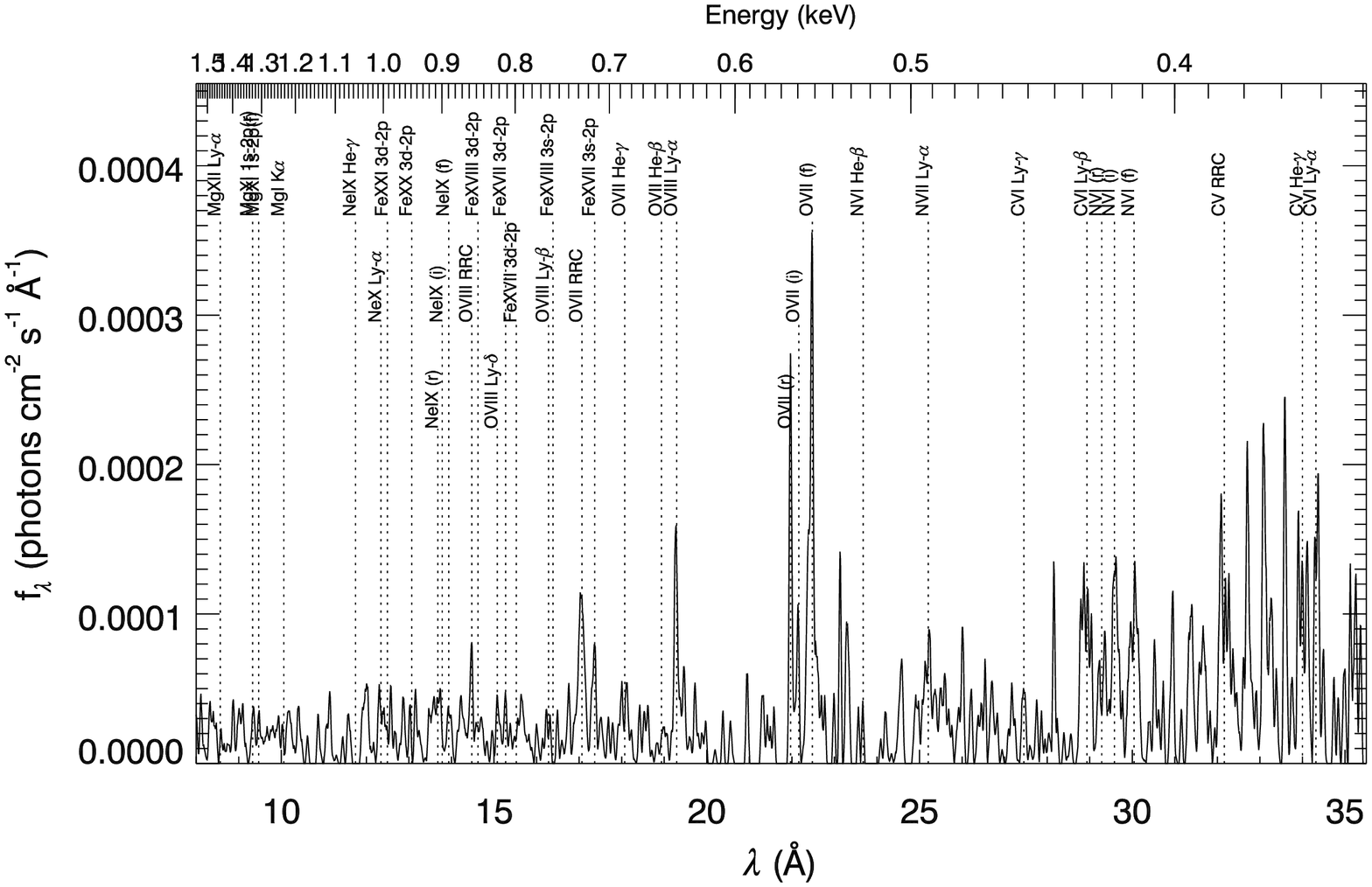}
\caption{\footnotesize{High resolution spectrum of Mrk\,573 obtained with RGS/\emph{XMM-Newton} instrument.}
\label{fig:RGSspec}}
\end{center}
\end{figure*}

\subsection{RGS/XMM--Newton high resolution spectra}

Soft X-ray emission in Seyfert galaxies has been proven to consist of  a plethora of emission lines plus a 
small fraction of continuum emission that can be described with a single flat power-law 
\citep{Guainazzi08} with a fixed spectral index of $\rm{\Gamma = 1}$. 
We obtained the emission line fluxes of the central 30 arcsec 
region (note that this  includes the nucleus and circumnuclear emission) using the RGS/{\it XMM--Newton} data.
We searched for the presence of 37 emission lines of C, O, N, Si, Mg and Fe species by fitting the spectra of the two
RGS cameras to Gaussian profiles together with a continuum. We used Cash statistic for this purposes.

The triplet fits were performed keeping the relative  
distance between centroids in energy and the centroid energy was left as a free parameter. A line was considered detected when the
flux was higher than 0 at the 1${\sf \sigma}$ level. 
The resulting RGS spectrum and detected emission lines are presented in Figure \ref{fig:RGSspec} and 
 Table \ref{tab:RGS}, respectively. All energy centroids are consistent with the laboratory value given the error bars.
 \citet{Guainazzi08} previously studied 
the RGS/\emph{XMM--Newton} spectra of Mrk\,573. Unfortunately, they only reported
some of the lines, all of them agreeing with our emission line fluxes. The most intense emission lines 
comprising the RGS/\emph{XMM--Newton} spectrum are: C VI Ly$\rm{\beta}$, O VII (r), O VII (f),
O VIII Ly$\rm{\alpha}$, O VII H$\rm{\gamma}$, O VII RRC, Fe XVII 3d-2p, and Ne IX (r).

\subsection{EPIC/XMM--Newton low resolution spectra}\label{sec:epicspec}


\begin{figure}[!t]
\begin{center}
\includegraphics[height=1.0\columnwidth,angle=270]{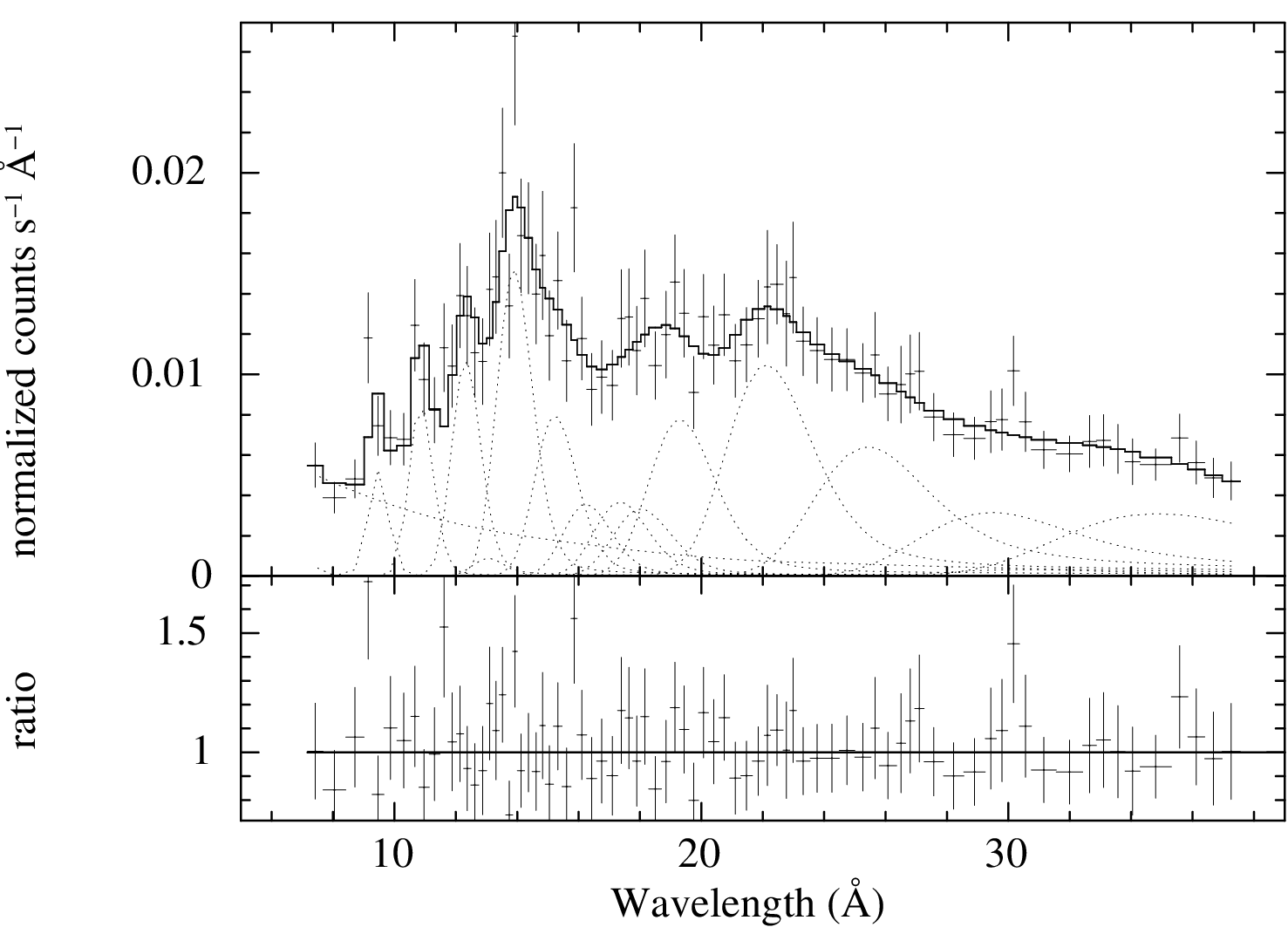}
\caption{\footnotesize{Spectral fits (top panel) and residuals (bottom panel) for the nuclear spectrum of Mrk 573
using low resolution EPIC pn/{\it XMM-Newton} data. Dotted-lines show the Gaussian components used to fit the spectrum.}}
\label{fig:xspecXMM}
\end{center}
\end{figure}

The fit of EPIC/\emph{XMM--Newton} data with a thermal model produces poor results below 2 keV 
($\sf{\chi^{2}}$ $\sim$16). 
Instead, a model composed of multiple emission lines was tried.
Taking advantage of the RGS/\emph{XMM--Newton} fit, 
we imposed that the intensity of the lines in the low resolution spectra fit do not exceed 
the RGS measurements. This is acceptable 
because the cross-calibrations between 
EPIC and RGS instruments shows a normalization constant in the range of 0.9 to 1.0 \citep[see][]{Plucinsky08}. 
 The assumed Gaussian width is 100 eV.
Note that for EPIC (and also ACIS/\emph{Chandra}) data we 
do not question the existence of the emission lines detected on the RGS/\emph{XMM-Newton} but 
we use them as a template. Triplets were fitted using the total flux 
of all components of the He-like lines O VII, N VI, and Ne IX. 
The continuum emission was fitted to a power-law to be consistent with the high spectral resolution analysis.
However, this fit has poor statistics ($\rm{\chi^{2}_{r} > 2}$). 
Five lines were added at energies above 0.95~keV in order to achieve an acceptable fit 
($\rm{\chi^{2}_{r} = 0.8}$): 
 Fe\,XX at  0.97~keV ($\rm{\chi_{r}^{2}}$=5.64), 
Ne\,X~Ly$\rm{\alpha}$ at 1.02 keV ($\rm{\chi_{r}^{2}}$=5.62), 
Ne\,IX He $\rm{\delta}$ at 1.16 keV ($\rm{\chi_{r}^{2}}$=3.72), 
Mg\,XI triplet at $\rm{\sim}$1.33 keV ($\rm{\chi_{r}^{2}}$=0.82),
and Si~XIII triplet at  1.84~keV($\rm{\chi_{r}^{2}}$=0.80).
The final fit is shown in Figure \ref{fig:xspecXMM}. 
The low spectral resolution EPIC/\emph{XMM--Newton} spectrum shows the following intense
emission lines: C\,V~He$\rm{\gamma}$, C\,VI~Ly$\rm{\beta}$, N\,VII~Ly$\rm{\alpha}$, O\,VII triplet,
O\,VIII Ly$\rm{\alpha}$, O VII\,He$\rm{\gamma}$,  O\,VII RRC, Fe\,XVII 3d-2p, and Ne\,IX triplet, 
Ne\,X~Ly$\rm{\alpha}$, Ne\,IX~He $\rm{\delta}$, and Mg~XI triplet. 
 In the best-fit model the flux of the OVIII RRC feature 
appears negligible and the adjacent line Fe\,XVII 3d2p is present. This is in contrast to
what happens in the ACIS/\emph{Chandra} nuclear spectrum (see Section \ref{sec:chandraspec} and Table \ref{tab:low}). In order to 
check the compatibility of the results of both spectra, we have imposed a zero intensity to the line Fe\,XVII 3d2p 
finding a good fit. In this case, the OVIII RRC feature shows a similar flux compared to the flux reported using the \emph{Chandra} nuclear 
spectrum. It is very likely that both features share the flux as measured in the RGS/\emph{XMM-Newton} spectrum, although
they cannot be distinguished in the low resolution spectra. 
The best-fit model has an 
absorbed 0.5-2.0 keV flux of 
F$\rm{_{0.5-2.0~keV}=}$3.38 (3.37-3.83) $\rm{\times 10^{-13}erg~s^{-1}~cm^{-2}}$, corresponding to an
unabsorbed rest frame luminosity of L$\rm{_{0.5-2.0~keV}=}$2.0 (1.9-2.2) $\rm{\times 10^{41}~erg~s^{-1}}$.
We have only included Galactic absorption in our model, which corresponds to $N_H = 2.52\times 10^{20}$.



\subsection{ACIS/Chandra spectra}\label{sec:chandraspec}

We cannot separate the contribution of NW and SE regions to the RGS or EPIC/\emph{XMM--Newton} spectra.
However, this is possible by using the 
lower spectral resolution but better spatial resolution of ACIS/{\it Chandra} data. 

\begin{figure}[!t]
\begin{center}
\includegraphics[width=1.0\columnwidth]{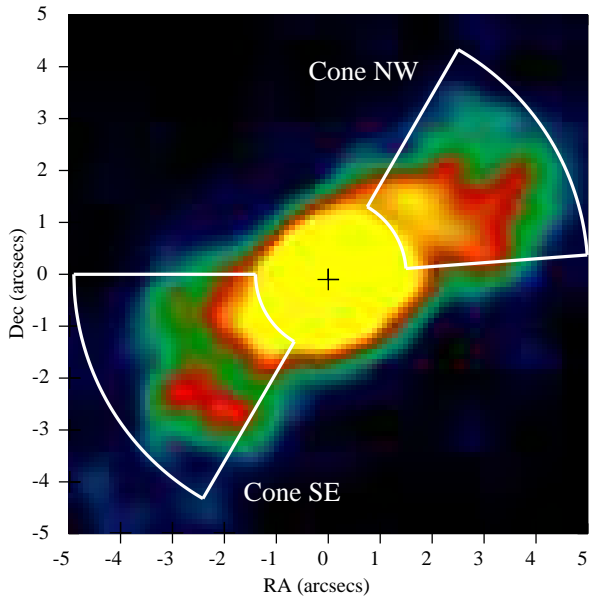}
\caption{\footnotesize{Soft (0.2-2 keV) X-ray {\it Chandra} 
image with the two extracted regions (NW and SE) overplotted.}
\label{fig:overcones}}
\end{center}
\end{figure}

We extracted spectra from the nucleus ($R=1$ arcsec)  and two conical regions coincident
with the extension of the emission, as seen in Figure \ref{fig:overcones}. All extraction regions were 
centered at (RA, Dec)=01:43:57.78,+02:20:59.4.
For the conical regions, we used an annulus, centered at the same position than the nucleus,
 with an inner and outer radius of 1.5 and  
5 arcsec, respectively. The cones are defined by an opening angle of 60${\sf^{o}}$ 
centered at $\mathrm{PA}=325^\circ$ (cone NW) and at $\mathrm{PA}=145^\circ$ (cone SE). 

The study of the emission above 2 keV is beyond  the scope of this paper since 
the nucleus dominates there and it has been already well studied before
\citep{Guainazzi05}. Therefore, channels above $\rm{\sim}$2~keV were ignored in the 
spectral fit. Again, the thermal model gives a poor fit with some residuals below 2 keV
($\sf{\chi^{2}}$ $\sim$3 for both cone regions). 

\begin{figure}[!t]
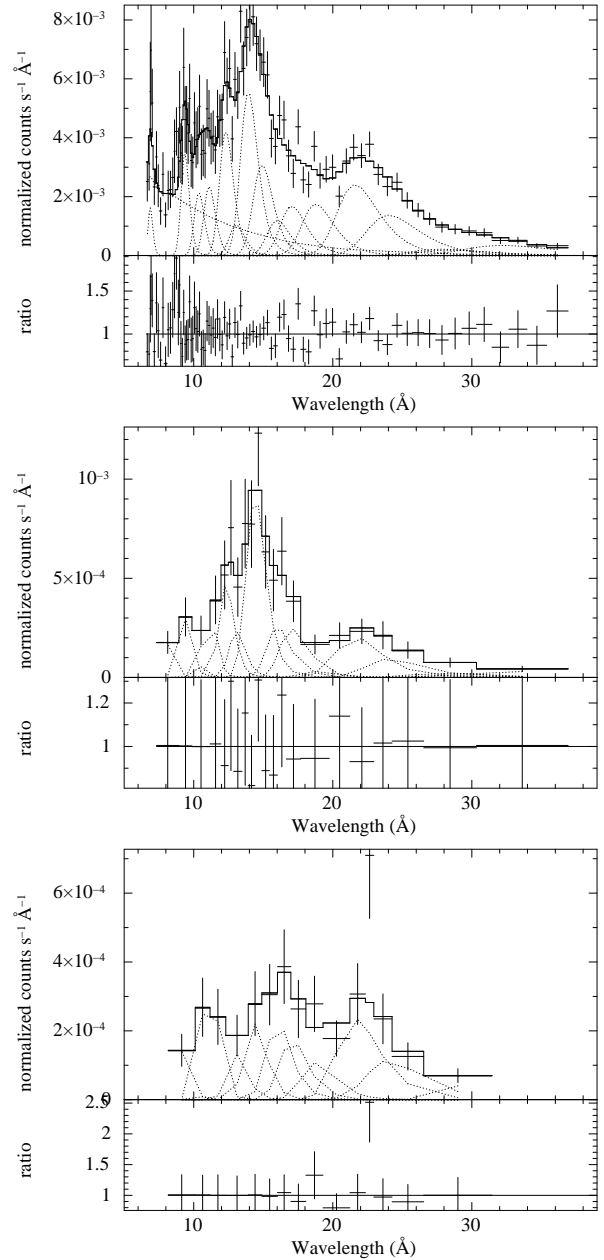

\begin{center}
\includegraphics[height=0.9\columnwidth,angle=-90]{Mrk573_nucleus_data_new.ps}\\
\includegraphics[height=0.9\columnwidth,angle=-90]{Mrk573_cono1_data_new.ps}\\
\includegraphics[height=0.9\columnwidth,angle=-90]{Mrk573_cono3_data_new.ps}
\caption{\footnotesize{Spectral fits ( top panels) and residuals ( bottom panels) 
using low resolution ACIS/\emph{Chandra} spectra of the 
three regions: Nucleus (top), cone NW (middle) and cone SE (bottom).}}
\label{fig:xspecsdata}
\end{center}
\end{figure}

We used the same model reported in the RGS/{\it XMM--Newton} spectrum assuming 
the emission line fluxes found in that case, 
as upper limits to fit the nucleus, and the NW and SE cone-like structures (the same as
the EPIC/{\it XMM--Newton} data mentioned in the previous section). 
Moreover, the power-law component was removed in the cone-like structures
because we expect this component to be detectable only in the nuclear region.
 The assumed Gaussian width is 100 eV.
The spectra are shown in Figure \ref{fig:xspecsdata} and final
emission line fluxes are reported in Table  \ref{tab:low}.

Most of the emission lines in the EPIC/\emph{XMM--Newton} spectrum are also 
detected in the nuclear spectrum using \emph{Chandra} data (CV H$\rm{\gamma}$, N VII Ly$\rm{\alpha}$, O VII triplet,
O VIII Ly$\rm{\alpha}$, Ne IX triplet, Ne X 
Ly$\rm{\alpha}$, and Mg XI triplet). Five of them were present only in the EPIC/\emph{XMM--Newton} spectrum 
(C VI Ly$\rm{\alpha}$, O VII H$\rm{\gamma}$, O VII RRC, Fe XVII 3d-2p, and Ne IX He$\rm{\delta}$). 
We note that the flux 
measurements in the ACIS/Chandra spectrum are compatible with those of the EPIC/XMM-Newton
including cases where only upper limits can be estimated in one of the spectra.
In contrast with the EPIC/{\it XMM--Newton} spectrum, the Ne\,IX~He$\rm{\delta}$ line at 1.16 keV was not needed,
instead two more lines were added  (Ne\,IX~He$\rm{\gamma}$ and Ne\,X~Ly$\rm{\gamma}$ at 1.13 and 1.22 keV, 
respectively) in order to get the best fit ($\sf{\chi^{2}}$=1.1). Note that in the case of Ne\,X~Ly$\rm{\gamma}$
there are several transitions of Fe~XX that could be contributing to this line. The inclusion of these 
two lines (i.e. Ne\,IX~He$\rm{\gamma}$ and Ne\,X~Ly$\rm{\gamma}$) instead of the Ne\,IX~He$\rm{\delta}$ line
also provides an aceptable fit to the EPIC/{\it XMM--Newton} spectrum. The line intensities are
reported in Table \ref{tab:low} (within brackets). The final fit is equaly acceptable. 
A similar case is affecting the OVIII RRC and the Fe\,XVII 3d-2p lines (see Section \ref{sec:epicspec}).

\begin{table*}
\begin{center}                          
\caption{Measured fluxes for EPIC/{\it XMM-Newton} and ACIS/{\it Chandra} spectra}
\begin{tabular}{l c c c c c c c c }   \\     
\hline\hline                 
          Line     & $\lambda_{rest}$ & Energy &  \multicolumn{1}{c}{XMM-Newton}	    &  	    \multicolumn{3}{c}{Chandra}	    	     \\ \cline{5-7} 		     
                   &                  &        &				            &   \multicolumn{1}{c}{Nucleus} & \multicolumn{1}{c}{Cone NW}  &  \multicolumn{1}{c}{Cone SE}     \\
                   & (\AA)            & (keV)  &	($10^{-5}$)$^a$           &  ($10^{-5}$)$^a$           & ($10^{-6}$)$^a$      &  ($10^{-6}$)$^a$     \\
 \hline
Norm (pw)	          &                   &         & $3.1_{2.4}^{3.7}$       & $2.5_{2.2}^{2.9}$      &  		...         &	     ...		\\
${\sf	   CV~He\gamma}$  & 33.43             & 0.371   & $18.6_{14.8}^{22.0}$	  & $12.1_{6.9}^{16.6}$	   &  $12.6_{0.0}^{22.5}$   &  $27.9_{2.0}^{53.8}$     \\
${\sf NVI~triplet}$       & 28.79/29.08/29.53 & 0.426   & $0.9_{0.0}^{3.6}$	  & $2.5_{0.0}^{7.1}$	   &  $1.5_{0.0}^{13.1}$    &	    ... 		\\
${\sf	  CVI~Ly\beta}$   & 28.47             & 0.436   & $7.4_{1.7}^{9.4}$	  & $2.3_{0.0}^{5.6}$	   &  $2.6_{0.0}^{10.6}$    &	    ... 		\\
${\sf	 NVII~Ly\alpha}$  & 24.78             & 0.500   & $7.0_{5.7}^{8.3}$	  & $5.3_{4.4}^{6.6}$	   &  $3.7_{0.0}^{ 7.6}$    &  $4.5_{1.0}^{7.9}$     \\
${\sf OVII~triplet}$      & 21.60/21.80/22.10 & 0.569   & $8.7_{7.5}^{9.9}$	  & $7.8_{6.4}^{9.2}$	   &  $6.9_{5.0}^{10.9}$    &  $7.7_{3.2}^{12.2}$     \\
${\sf	OVIII~Ly\alpha}$  & 18.97             & 0.654   & $3.5_{2.4}^{4.7}$	  & $2.3_{1.4}^{2.9}$	   &  $0.5_{0.0}^{2.2}$     &  $1.5_{0.0}^{3.7}$	   \\
${\sf	 OVII~He\gamma}$  & 17.77             & 0.698   & $1.2_{0.2}^{2.8}$	  & $0.0_{0.0}^{1.4}$	   &  $0.0_{0.0}^{2.20}$    &	   ...  		\\
${\sf FeXVII~3s2}$        & 17.08             & 0.726   & $1.2_{0.0}^{2.2}$	  & $1.6_{0.1}^{2.2}$	   &  $2.4_{0.0}^{3.67}$    &  $1.7_{0.0}^{4.0}$	   \\
${\sf	OVII~RRC}$        & 16.77             & 0.775   & $1.0_{0.1}^{1.9}$	  & $0.8_{0.0}^{1.5}$	   &  $1.8_{0.0}^{3.64}$    &  $1.6_{0.0}^{3.3}$	   \\
${\sf FeXVII~3d2p}$       & 15.01             & 0.826   & $1.7_{0.7}^{2.2}$	  & $0.0_{0.0}^{1.7}$	   &  $0.0_{0.0}^{2.89}$    &	   ...  		\\
${\sf  OVIII~RRC}$        & 14.23             & 0.871   & [$1.6_{1.0}^{2.0}$]\,$^b$ & $1.5_{0.0}^{2.6}$	   &  $4.08_{0.43}^{5.15}$  &  $1.0_{0.2}^{1.8}$     \\
${\sf NeIX~triplet}$      & 13.45/13.55/13.70 & 0.905   & $2.5_{1.7}^{3.0}$	  & $2.2_{1.6}^{2.6}$	   &  $0.0_{0.0}^{3.1}$     &	    ... 		\\
${\sf FeXX~3d2p}$         & 12.85             & 0.965   & $0.1_{0.0}^{0.6}$	  & $0.32_{0.02}^{0.70}$   &  $0.76_{0.01}^{1.65}$  &  $0.6_{0.1}^{1.0}$   \\
${\sf NeX~Ly\alpha}$      & 12.13             & 1.022   & $1.3_{1.0}^{1.6}$	  & $1.1_{0.9}^{1.3}$	   &  $1.3_{0.4}^{2.1}$     &				\\
${\sf NeIX~He\gamma}$     & 10.97             & 1.130   & [$0.7_{0.4}^{0.8}$]\,$^c$ & $0.44_{0.29}^{0.59}$   &  $0.7_{0.0}^{1.3}$     &  $0.9_{0.4}^{1.3}$   \\
${\sf NeIX~He\delta}$     & 10.69             & 1.160   & $0.8_{0.6}^{1.0}$	  &	   ...  	   &		...	    &	    ... 		\\
${\sf NeX~Ly\beta}$       & 10.16             & 1.220   & [$0.3_{0.1}^{0.5}$]\,$^c$ & $0.33_{0.20}^{0.46}$   &  $0.13_{0.0}^{0.6}$    &  $0.02_{0.0}^{0.5}$	    \\
${\sf MgXI~triplet}$      & 9.17/9.31         & 1.352   & $0.4_{0.2}^{0.6}$	  & $0.41_{0.30}^{0.52}$   &  $0.5_{0.2}^{0.8}$     &  $0.45_{0.2}^{0.7}$  \\
${\sf SiXIII~triplet}$    & 6.65/6.74         & 1.840   & $0.5_{0.0}^{2.8}$	  & $0.11_{0.02}^{0.19}$   &  $2.0_{0.9}^{3.1}$     &	    ... 		\\
\hline					   
\end{tabular} \\
\tablenotemark{a}{Units are $\mathrm{ph}\, \mathrm{cm}^{-2}\, \mathrm{s}^{-1}$.}							        				 
\tablenotemark{b}{This value was obtained by setting the nearby line FeXVII to zero.}
\tablenotemark{c}{These values were computed after including 
Ne\,IX~$\mathrm{He}\gamma$ and Ne\,X~$\mathrm{Ly}\beta$, instead of NeIX~$\mathrm{He}\delta$.}										        		
\label{tab:low}      
\end{center}
\end{table*}

The emission lines in the NW and SE cones are about a factor of 10 or more, fainter than those
in the nuclear region using \emph{Chandra} data. In the NW cone we have found the O VII triplet,
O VIII RRC, Fe\,XX~3d2p, Ne\,X~Ly$\rm{\alpha}$, Mg XI triplet and Si XIII triplet. 
A similar result is found in the SE cone with the detection of 
C V H$\rm{\gamma}$, N VII Ly$\rm{\alpha}$, O VII triplet,
O VIII RRC, Ne IX triplet, Fe XX, Ne IX He$\rm{\delta}$, 
and Mg XI triplet. All of these lines were detected in the ACIS/\emph{Chandra} 
nuclear region, except O VIII RRC.

\begin{figure}[!t]
\begin{center}
\includegraphics[width=1.0\columnwidth]{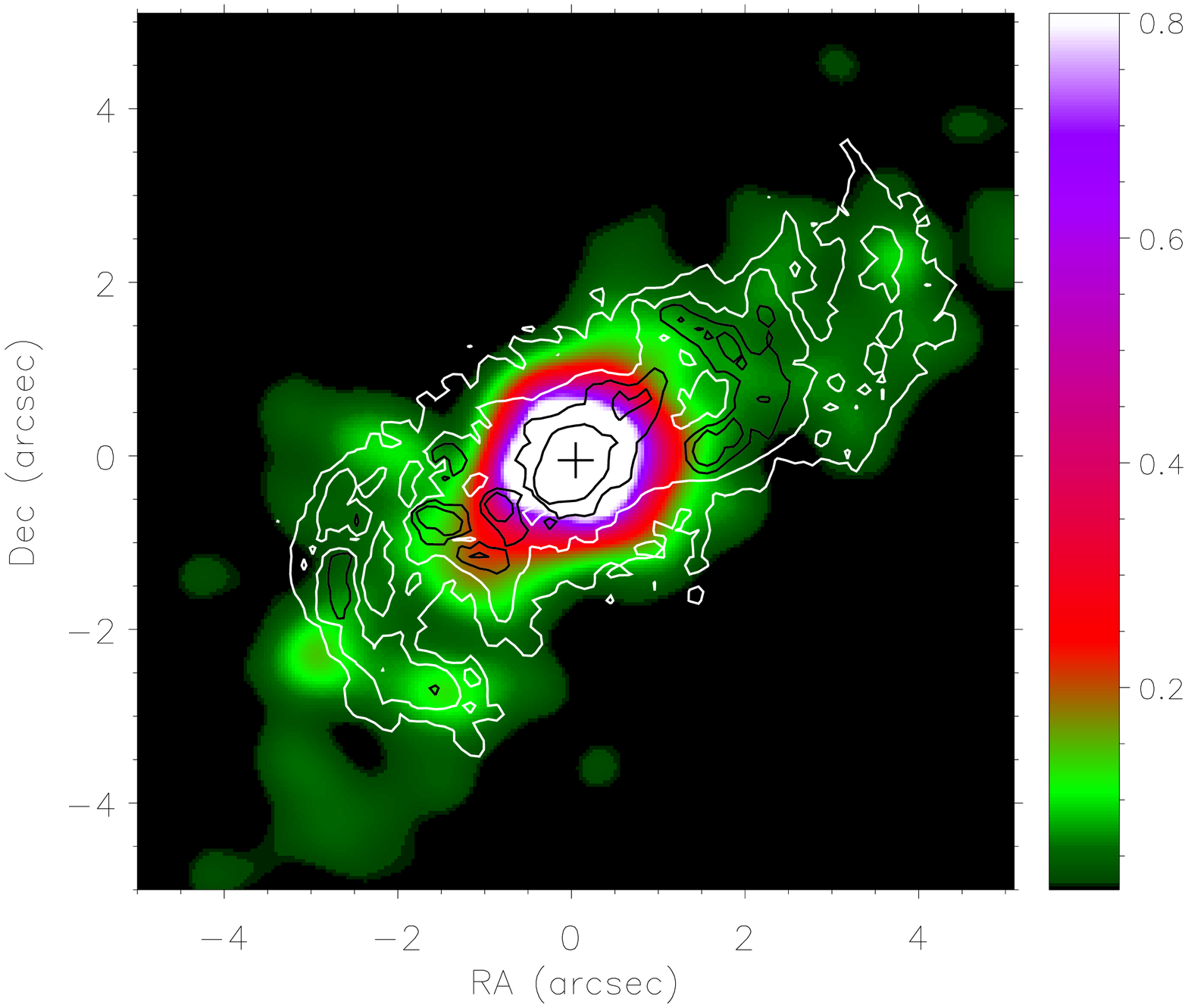}
\includegraphics[width=1.0\columnwidth]{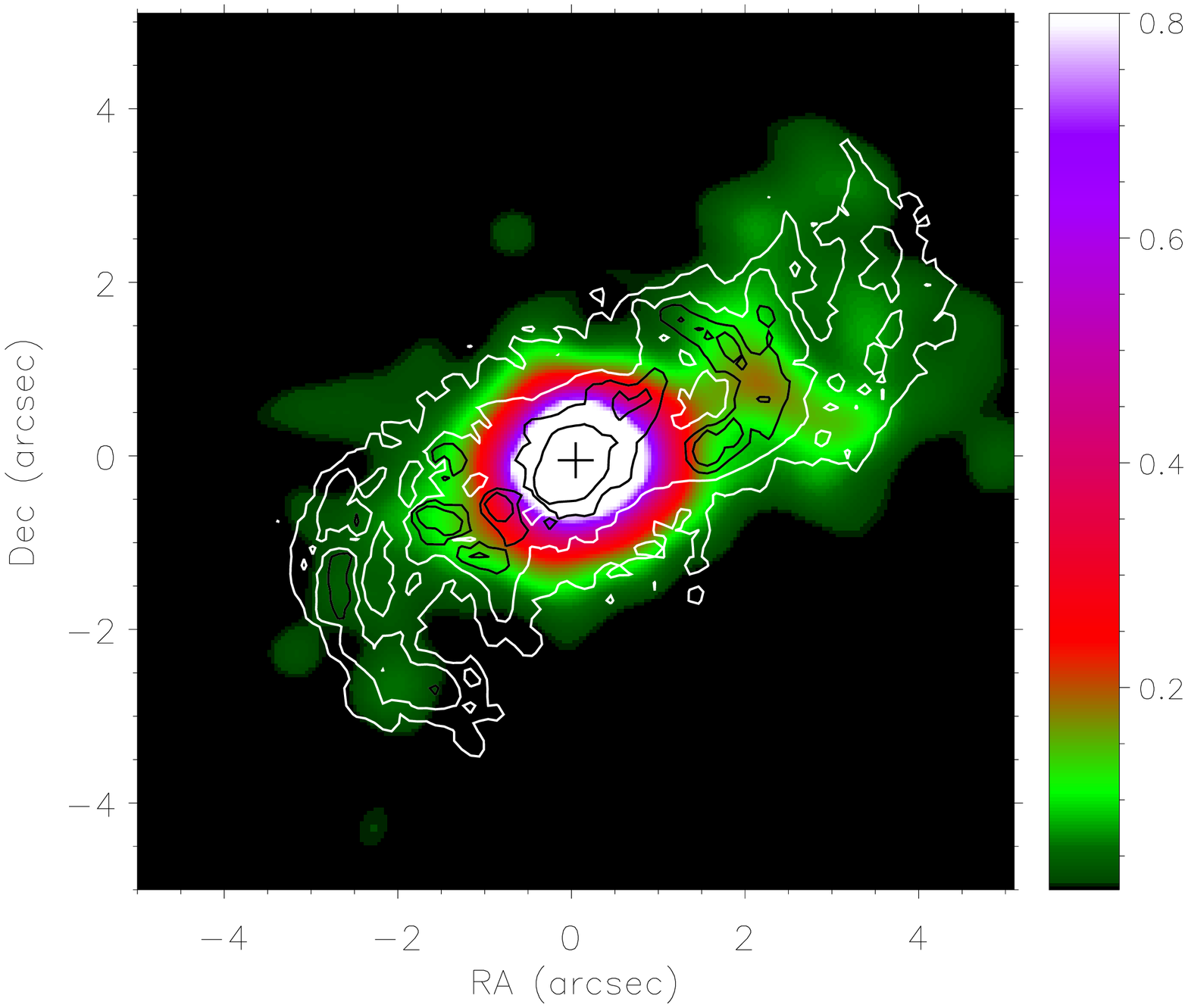}
\caption{\footnotesize{Images centred at O VII triplet (top) and O VIII RRC (bottom) lines. [O III] contours are overlaid.}
\label{fig:imaxraylines}}
\end{center}
\end{figure}

\begin{figure}[!t]
\begin{center}
\includegraphics[width=1.0\columnwidth]{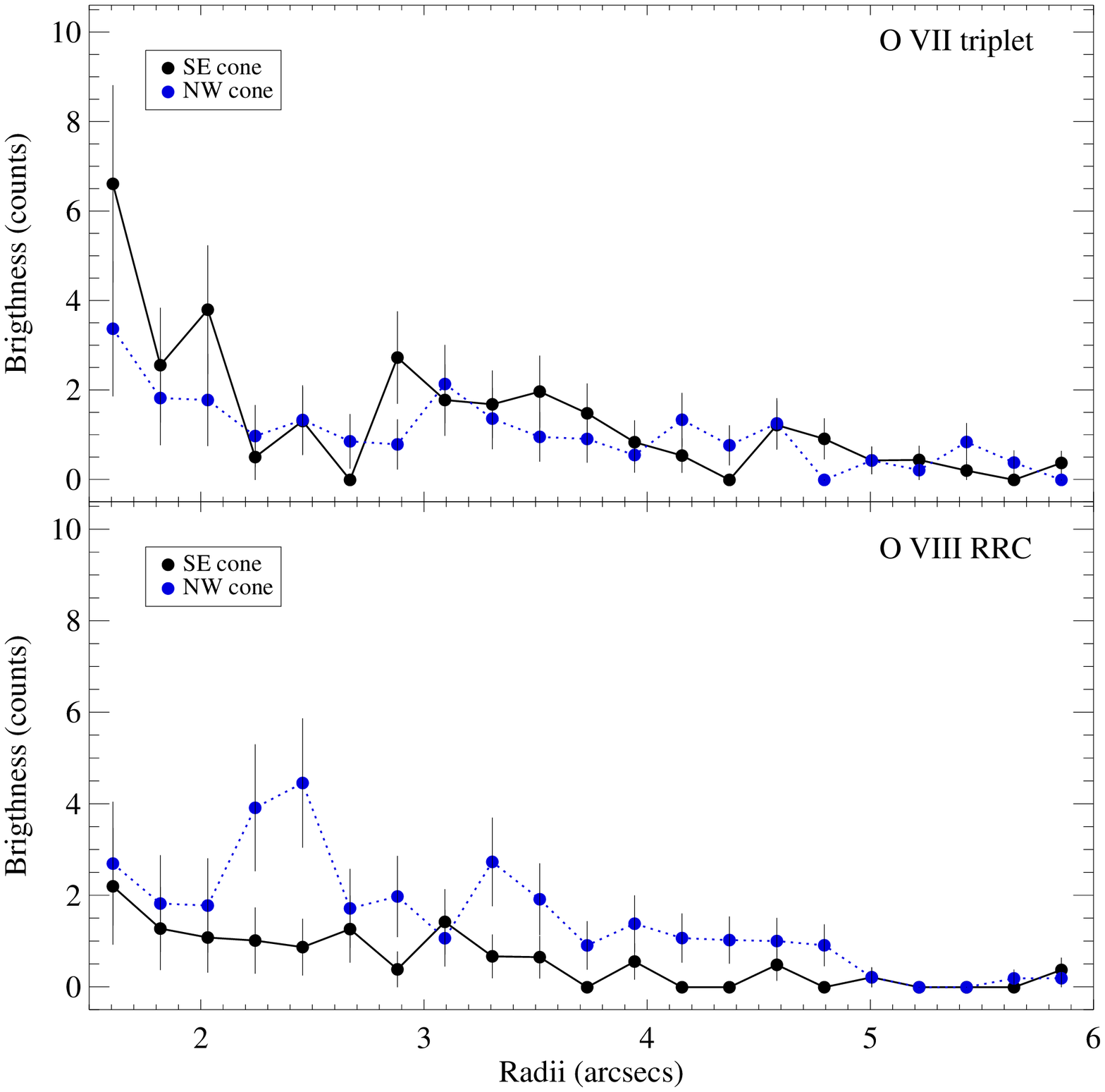}
\caption{\footnotesize{Radial profile of the images centred at O VII triplet (top) and O VIII RRC (bottom) lines. 
Blue-dotted line and blue dots correspond to the NW cone while black-continous line and black dots correspond to the SE cone.}
\label{fig:radial_OVIII}}
\end{center}
\end{figure}

At first glance the spectra of the cones look rather different (see Figure \ref{fig:xspecsdata}), 
the NW cone spectrum shows an enhancement around 14\AA (0.89 keV), which is not present in that of
the SE cone. To illustrate this effect we have constructed 
two images centered at 0.5-0.7 keV and 0.8-1.0 keV, respectively (see Figure \ref{fig:imaxraylines}).
These spectral intervals are dominated by the emission of the O~VII triplet at 0.569 keV and O~VIII RRC at 0.871 keV, 
respectively (see Tab.  \ref{tab:low}). The O VII triplet image shows a morphology similar to that found in the soft band 
($\rm{<}$ 2 keV, see Figure \ref{fig:X_OIII}). 
However, the O VIII RRC image shows a bright region to the NW of the nucleus. This can also be noticed in Figure 
\ref{fig:radial_OVIII} where the radial profiles of the O~VII triplet and O~VIII RRC images are plotted. 
The NW and SE profiles appear consistent in the case of the O VII triplet, whereas there is a clear bump 
at 2.5"  in the NW side of the O~VIII~RRC profile compared to the SE one.
Despite the limited signal-to-noise ratio of the spectra from the cones, it seems clear that
the  O VIII RRC is more important in the NW cone than in the SE cone 
(see Figs. \ref{fig:xspecsdata} and \ref{fig:radial_OVIII}).

\section{Origin of the soft X-ray emission}\label{sec:origin}

As already said, the soft X-ray spectrum of Mrk\,573 is dominated  by line emission, as evident from  the 
RGS/{\it XMM--Newton} data (Figure \ref{fig:RGSspec}). 
Emission lines from H-like and He-like C, N, O, and Ne and Fe L-shell 
emission line Fe XVII dominate the spectrum, which also includes strong narrow 
RRC O VII and O VIII lines. The RRC features  of these highly ionized species are produced when electrons recombine directly 
to the ground state. They are broad features for hot, collisionally ionized plasma, but 
are narrow, prominent features for photoionized plasma arising in low-temperature material \citep{Liedahl96,Liedahl99}.
\citet{Guainazzi07} measured in the Mrk~573 RGS spectrum the width of the RRC lines O VII and O VIII as $\rm{\sigma(O~ VII)=3.5_{-3.3}^{+5.3}}$ eV 
and $\rm{\sigma(O~ VIII)=4.8\pm3.1}$ eV, respectively. These values indicates a relatively cool photoionized plasma.

Relative emission strength ratios of the He-like $\rm{1s2p~^{1}P_{1}\rightarrow~1s^{2}~^{1}S_{0}}$ 
(r, resonance), $\rm{1s2p~^{3}P_{2}\rightarrow~1s^{2}~^{1}S_{0}}$,  $\rm{1s2p~^{3}P_{1}\rightarrow~1s^{2}~^{1}S_{0}}$ 
(i, blended inter-combinations)
and $\rm{1s2s~^{3}S_{1}\rightarrow~1s^{2}~^{1}S_{0}}$ (f, forbidden) transitions can discriminate between 
photoionization, collisional excitation, or hybrid environment \citep{Porquet00,Bautista00,Kahn02}. 
A weak resonance ($r$) line, compared to the forbidden ($f$) and intercombination ($i$) lines, corresponds to 
a pure photoionized plasma. A commonly used line ratio is defined as $G=(f+i)/r$. 
For Mrk\,573 we measured a value of $G = 1.4 (0.3-12)$ for the ion O~VII (see Tab. \ref{tab:RGS} and Figure \ref{fig:RGSspec}). 
This value is lower than that expected for a pure photoionized plasma ($G > 4$), but higher than that of 
collisional ionization. Such a value of G would indicate a hybrid plasma where collisional processes are not negligible 
\citep{Porquet00}, although the error bars do not allow a secure result. 
We could only obtain upper limits in the intensity of two of the lines needed to compute the $G$ ratio corresponding to the
other ions (Ne~IX and N~VI) with detected He-like triplet transitions. The lower limit of G obtained using 
Ne~IX ($\rm{(f+i)/r\gtrsim 2.7}$) is not conclusive although also points to an hybrib plasma. 
However, the use of $G$ ratios has been questioned by several authors 
\citep[e.g.][]{Kinkhabwala02,Porter07}. 
An important contribution of photoexcitation would rise the intensity of the resonance transition, which yields a 
decrease of the $G$ ratio relative to the pure photoionization case.  \citet{Kinkhabwala02} proposed the use of 
the ratio of higher order transitions to the forbidden triplet component in He-like ions as a good discriminant 
between photoexcitation and collisional ionization. 
 \citet[][]{Kinkhabwala02} predicted a value of $\rm{F(O~VII~He\gamma)/F(O~VII~(He \alpha (f))}$ of 0.017 for 
collisional ionization.
Thus, an excess of He$\gamma$ line relative to the forbidden transition, as measured in 
Mrk~573, $\rm{F(O~VII~He\gamma)/F(O~VII~(He \alpha (f))}$ $\rm{=0.31\, (0.03-1.4)}$  
(see Tab. \ref{tab:RGS}), indicates the importance of photoexcitation.

The intensities of the emission lines Fe~XVII L (3d-2p) and Fe~XVII K (3s-2p) are 
comparable in the case of Mrk~573 (see Tab. \ref{tab:RGS}), which could be explained as an intermediate temperature plasma 
($10\,\mathrm{eV}<\mathrm{kT}<500\,\mathrm{eV}$), or under hybrid conditions combining collisional and photoionization equilibrium
conditions \citep{Liedahl90}. Similarly, \citet{Sako00} argued that the dominance of Fe~XVII L lines 
can only be explained if photoexcitation by the nuclear radiation plays an important role, consistent with our suggestion throughout 
the G ratio.

From above, we conclude that the emission line diagnostics seems to indicate that the soft-X ray 
spectrum of Mrk~573 as seen by RGS/\emph{XMM-Newton} data can be produced by a plasma dominated by photoionization and photoexcitation. 
Nevertheless, we cannot rule out some contribution of collisional excitation, based primarily on 
the presence of the Fe~XVII 3d-2p transition.    
Unfortunately, the limited count level of the spectra does not permit us to give secure diagnostics.
Detailed optical spectroscopic studies by \citet{Ferruit99} found that ionizing photons 
originating in the central source are not sufficient to explain the emission line luminosity. 
They suggest that fast shocks, associated with the jet/gas interactions, 
might contribute to the gas ionization.

The ACIS/\emph{Chandra} spectrum of the nuclear region (1 arcsec) 
can be reproduced using the template obtained from the RGS/\emph{XMM--Newton} analysis. 
The resulting fluxes are consistent with each other. Therefore, it seems that most 
of the emission line fluxes measured in the RGS spectra come from the inner central 2 arcsec region and 
is consistent with the photoionization and photoexcitation dominated plasma scenario.

NW cone emission differs from that of the SW cone and the nucleus. 
The O VIII~RRC/O~VII triplet ratio found in the NW cone ($\rm{F(O VIII~RRC)/F(O~VII(r,i,f))}$ $\rm{=0.6_{0.1}^{1.1}}$) is nominally higher than 
in the SE cone ($\rm{0.13_{0.02}^{0.56}}$) and in the nucleus ($\rm{<0.19}$), although consistent within the statistical uncertainties.
According to \citet{Kinkhabwala02} radiative decay following photoexcitation dominates the 
Seyfert 2 spectrum at low column densities, whereas recombination following photoionization dominates at
high column densities. Following their prescription, the RRC intensity will increase compared to
He-like triplet line when column density increases, which would imply a value of $\mathrm{N_H}$ 
larger for NW than for SE.
Alternatively, the RRC line could be produced by hot collisionally ionized plasma, although
the line widths reported by \citet{Guainazzi07} do not support this scenario. It is 
very unlikely that a broad feature contributes 
only in a few percentage of the total flux of source.  
Moreover, the radio maps of Mrk\,573 are not sensitive enough
to show  great detail concerning radio emission although two faint blobs are observed in the NW cone 
\citep{Falcke98}.
The low signal-to-noise ratio of the spectra from the cones prevents us from a more detailed study
of the line transitions in the extended emission.

\subsection{Comparison with photoionization models}

\begin{figure}
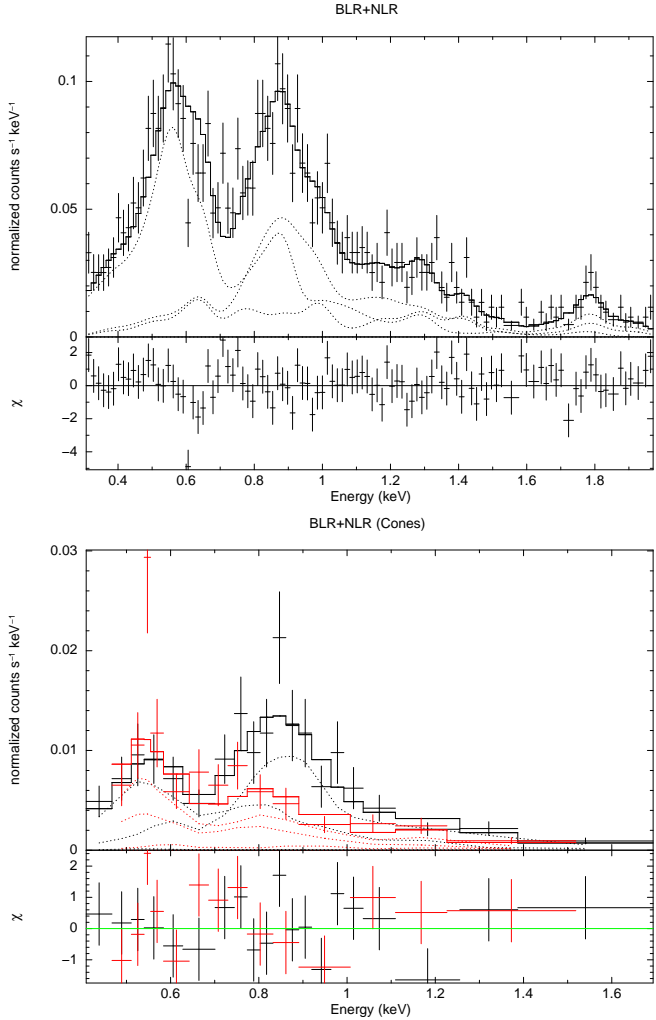

\begin{center}
\includegraphics[height=1.0\columnwidth,angle=-90]{BLRNLR-nucleus.ps}
\includegraphics[height=1.0\columnwidth,angle=-90]{Cones-BLRNLR.ps}
\caption{\footnotesize{(Top): ACIS/\emph{Chandra} spectral fit and residuals to BLR+NLR Cloudy model of the nuclear region. 
(Bottom):  Same for the cone-like regions. Red (black) crosses and red (black) continuous line are the NW (SE) cone spectrum and fit, respectively.}
\label{fig:spectrumCloudy}}
\end{center}
\end{figure}

\begin{figure}
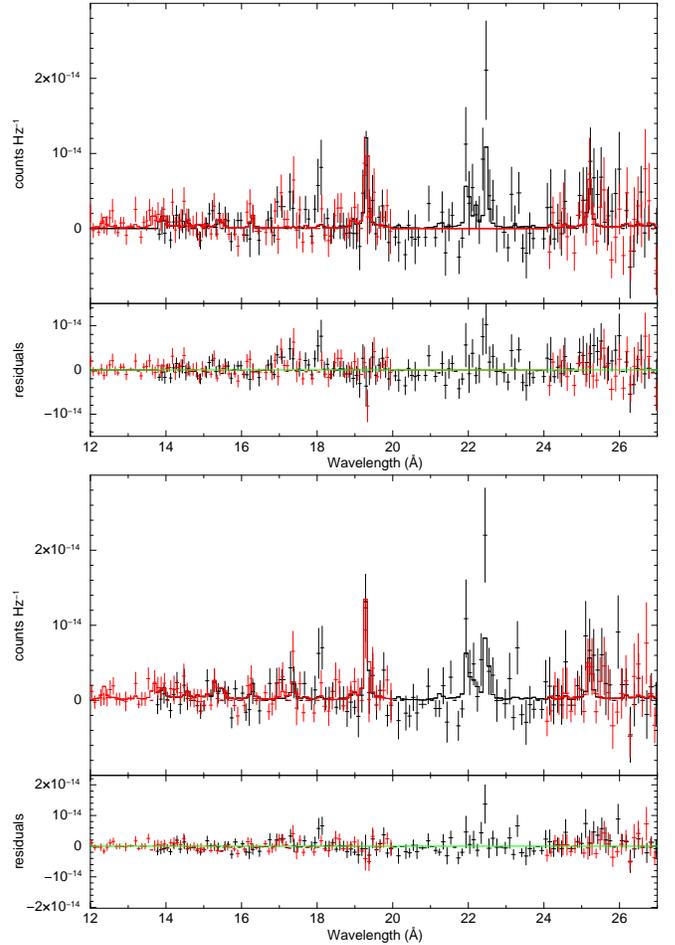

\begin{center}
\includegraphics[height=1.0\columnwidth,angle=-90]{MRK573_cloudy.ps}
\includegraphics[height=1.0\columnwidth,angle=-90]{MRK573_apec_cloudy.ps}
\caption{\footnotesize{(top): RGS/\emph{XMM-Newton} spectra fitted to the best fit found for ACIS/\emph{Chandra} 
spectrum of the nuclear region (see Figure \ref{fig:spectrumCloudy}). (bottom): Same than top figure but with the inclusion
of the thermal model APEC with kT=0.4$\rm{\pm 0.1}$ keV.}
\label{fig:spectrumCloudyRGS}}
\end{center}
\end{figure}

\begin{figure}
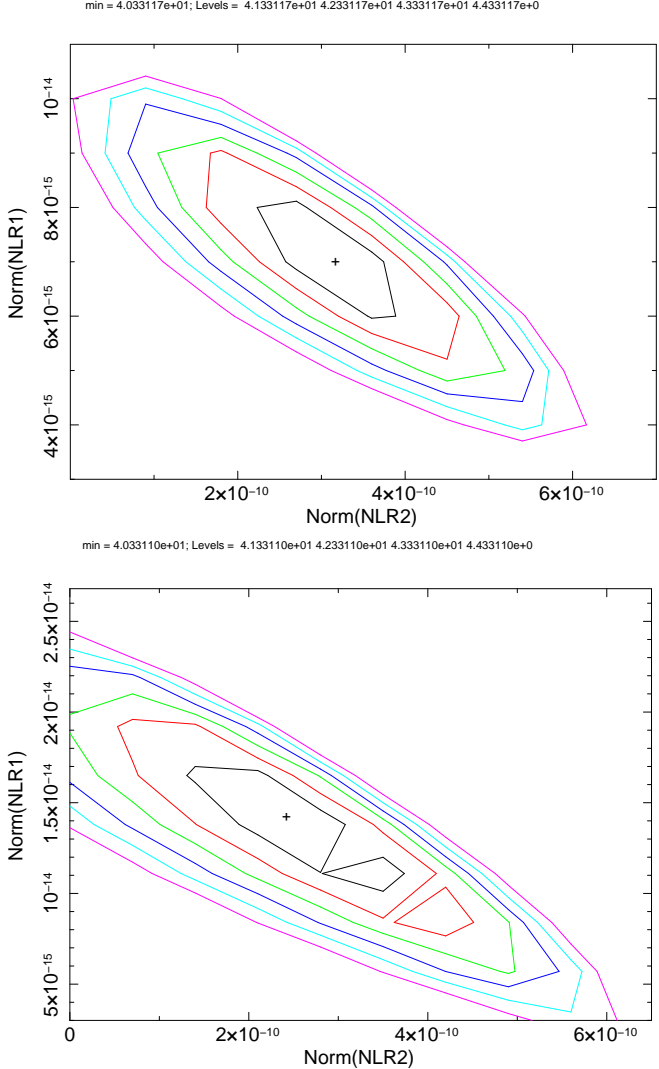

\begin{center}
\includegraphics[height=1.0\columnwidth,angle=-90]{contours_NW.ps}
\includegraphics[height=1.0\columnwidth,angle=-90]{contours_SE.ps}
\caption{\footnotesize{ 2-D iso-chi-squared flux contours of the normalization parameters of the 
best-fit Cloudy model of the reflected components of the NW (top) and SE (bottom) cones.}
\label{fig:isocontours}}
\end{center}
\end{figure}

Using version c08.00 of the Cloudy package \citep[last described by ][]{Ferland98}, we 
attempted to reproduce the observed spectra of the nuclear region and NE and 
SW cones seen in ACIS/\emph{Chandra} data.

In these Cloudy simulations we assumed the source of ionization to emit as 
a typical AGN continuum (we used the model AGN available in Cloudy) defined by 
a ``big bump'' of temperature $\mathrm{T} = 10^6 \rm{K}$, an X-ray to UV ratio 
$\rm{\alpha_{OX}=-1.15}$, plus a X-ray power-law of spectral index of 
$\rm{\alpha = -1.0}$. A plane--parallel geometry is assumed, with the slab depth controlled by 
the hydrogen column density  parameter ($\rm{N_H}$). 

Two grids of parameters were constructed to simulate the expected BLR 
and NLR conditions. For each of them, a grid of models was simulated by varying the
ionization parameter (U), the density of the material ($\mathrm{n_H}$), and the hydrogen 
column density ($\rm{N_H}$). 
U ranges from $\rm{10^{-3}}$ to $\rm{10^{3}}$, $\rm{N_H}$ 
ranges from $\rm{10^{20}\, cm^{-2}}$ to  $\rm{10^{24}\, cm^{-2}}$ 
and $\mathrm{n_H}$ from $\rm{10^{9}\, cm^{-3}}$ to $\rm{10^{11}\, cm^{-3}}$ for BLR conditions, 
and from $\rm{10^{2}\, cm^{-3}}$ to  $\rm{10^{4}\, cm^{-3}}$ for NLR conditions.

In both BLR or NLR models the outputs of each Cloudy simulation 
include a transmitted and a reflected emission line spectra. 
Under the Cloudy terminology ``reflected'' spectrum refers to the
emission escaping into the $2\pi$~sr subtended by the illuminated face towards the ionizing source and
by ``transmitted'' the emission escaping in the opposite direction.   
Each transmitted or reflected simulated emission line spectrum is imported 
as additive tables (``atable'') in XSPEC following \citet{Porter06} procedure.
All components are also absorbed through the Galactic value by using ``wabs'' on XSPEC.
In practice, any additive table is included in XSPEC as follows:  
$\mathrm{wabs[N_{H}(Gal)] (atable\{Cloudytable\}}$), where ``Cloudytable'' 
can be transmitted/reflected for BLR/NLR conditions. 

\begin{table*}
\begin{center}   
\caption{Best fit parameters to a Cloudy BLR+NLR model. Nucleus and Cones.}
\begin{tabular}{lcccc}
\hline \hline
                                 &  \multicolumn{2}{c}{Nucleus}                           &   \multicolumn{2}{c}{Cones (NW/SE)}   \\ 
Model param.			 &  BLR                      &  NLR                       & NLR 1                         & NLR 2                            \\ \hline
log(U)                           & $\rm{1.23_{1.18}^{1.27}}$ & $\rm{0.13_{0.11}^{0.16}}$  &  $\rm{0.9\pm0.3}$/$\rm{0.3\pm0.2}$  &  -3$^{*}$/-3$^{*}$       \\
log($n_H)$                       & $\rm{ 9.8_{9.7 }^{10.0}}$ & $\rm{3.14_{3.10}^{3.18}}$  &  3$^{*}$    &  3$^{*}$        \\
$\rm{log(N_H)}$ [$\mathrm{cm}^2$]&$\rm{20.6_{20.5 }^{20.7}}$ & $\rm{20.6_{20.5 }^{20.7}}$ &    20$^{*}$          &  20$^{*}$                 \\
Flux(0.5-2.0 keV) (trans.)        & $\rm{35.8\pm9.6}$           &   0.0 ($\rm{<6.6}$)           & 0.0 ($\rm{<5.3}$) / 9.3 ($\rm{<11.6}$)      & 3.4 ($\rm{<10.9}$) / 3.1 ($\rm{<9.3}$)	  \\
Flux(0.5-2.0 keV) (reflec.)       & $\rm{61.1\pm9.9}$           &  $\rm{92.6\pm3.7}$                      & 14.0 ($\rm{4.6-23.0}$) / 0.0 ($\rm{<14.5}$) & 3.8 ($\rm{<10.9}$) / 3.0 ($\rm{<9.3}$)	      \\
\hline
\end{tabular}
\tablecomments{NLR1 and NLR2 refer to high and low ionization phases for the cones
using ACIS/\emph{Chandra} data. Fluxes in units of $10^{-15}\mathrm{erg~cm^{-2}~s^{-1}}$.} 
\label{tab:cloudyfit}
\end{center}
\end{table*}

For the nuclear region, neither BLR nor NLR models produced any 
acceptable fit ($\rm{\chi^{2}_r=1.9}$ and $\rm{\chi^{2}_r=1.8}$, respectively). 
Qualitatively, the issue is that BLR and NLR models alone failed in
reproducing the relative strength between 0.55 keV (the O VII triplet) and 0.9 keV (the O VIII RRC and/or Ne IX triplet). 
We have found that the result of our Cloudy simulations do not change with the 
value of the volume density and column density in the range considered here. The models are
mostly sensitive to the value of the ionization parameter.
Besides the two simulations for BLR and NLR conditions (hereinafter called BLR and NLR models), 
we have also tried to model the spectrum of the nucleus as a combination of both BLR and NLR conditions 
(hereinafter BLR+NLR model). In the BLR+NLR model we use them all together,
being then two transmitted and two reflected emission line spectra. 
The combination of BLR+NLR model produced a good fit (see Figure 
\ref{fig:spectrumCloudy}) with $\rm{\chi^{2}_r=1.3}$. The parameters of the best fit are given in 
Table \ref{tab:cloudyfit}. 
The modelled nuclear spectrum is equally dominated by the BLR and NLR emission 
(BLR is $\rm{\sim}$54\% of the emission). We emphasize that in our simulation the use of a BLR 
component does not necessarily implies high density material, it rather accounts for 
ionization  conditions higher than  those of the NLR component. 

In principle we could also expect some contribution from collisionaly ionised plasma 
(see Section \ref{sec:origin}). However, we know that this contribution should be small 
as indicated by the G ratios, and discussed in previous section.
As an additional check, we used then this best-fit solution in order to reproduce 
the RGS/\emph{XMM-Newton} spectra. 
We froze all the parameters adding a constant to the fit since we were mainly interested to check
whether this model could reproduce the high resolution RGS spectra. 
We preferred  this approach of fitting first the low resolution spectrum and taking this as a starting 
point to model the high resolution spectra.
The final fit is given in Figure \ref{fig:spectrumCloudyRGS} (top) with 
a constant value of $\rm{1.2\pm0.2}$. It gives a good representation of the data although it fails in 
reproducing the Fe\,XVII emission lines. 
Actually the inclusion of a thermal model (APEC) 
with a kT=0.4$\rm{\pm0.2}$ keV reproduces better these features ($\rm{\Delta C\sim14}$, see Figure \ref{fig:spectrumCloudyRGS} bottom).
In the later case, it becomes very complex to discriminate between fits due 
to the low count level present in the high resolution spectrum. 
Based on this result, we went back to the low resolution data and try to fit the \emph{Chandra} 
nuclear region adding the APEC component. Although it does not give formally a better fit,
we obtained a fraction of the thermal component of $\rm{\sim}$ 6\% of the total nuclear flux. 
This results confirms that the thermally ionized component is present, 
although its contribution is small.

We have also modelled the extended emission using the grid of Cloudy simulations.  
In this case we have to include the contamination from the PSF wings of the 
nuclear source. For this, we have computed the expected fraction of the nuclear flux 
contributing to the cone-like regions as 
$F\rm{_{nuc2cone}} =$$Flux_{PSF}\mathrm{(cone)}$$/Flux_{PSF}$ $\mathrm{(nucleus)}$. 
Assuming a Gaussian profile for the \emph{Chandra} PSF with a FWHM of 1.2 arcsecs (see Figure \ref{fig:PSF}), 
the fraction of nuclear spectrum contributing to the cone is $F\rm{_{nuc2cone}}=0.006$. 
This factor is equivalent to an integrated flux of 
F(0.5-2.0 keV)= $\rm{5.06\times10^{-16}~erg~cm^{-2}s^{-1}}$, which corresponds 
to 2.5 and 3.5\% of the NW and SE cone fluxes, respectively.
In order to account for this nuclear contribution we have used the BLR+NLR best fit model found 
for the nucleus with all the parameters frozen, and scaled by the factor F$\rm{_{nuc2cone}}$. 
Initially we have tried to fit the cone emission by a single phase medium, using Cloudy models with NLR conditions and
both reflection and transmission components. Both NW and SE cones have been fitted simultaneously,
although the normalizations of the reflection and transmission components for the two cones are let
to vary independently.
The hydrogen column density and density of the material have been frozen at log($\rm{N_H}$)=20 and
log($\rm{n_{H}}$)=3 for simplicity. 
We recall that the simulated spectra by our Cloudy models are not sensitive to the value of 
these parameters within the explored range. 
The best fit show ionization parameters of log(U)=0.71$\rm{\pm}$0.07 and 
log(U)=0.13$\rm{\pm}$0.09 for NW and SE cones, respectively. Nevertheless, 
this fit failed to simultaneously reproduce the two spectra, being 
the statistics rather poor ($\rm{\chi^2_r=1.6}$). 
A much better fit is obtained ($\rm{\chi^2_r=1.1}$) if we add to this phase (hereinafter NLR1) 
a second one (hereinafter NLR2) with NLR conditions but a lower value of the ionization parameter U. 
We note that due to convergence problems within XSPEC we had to freeze the 
ionization parameter of the second phase. We checked that the best fit is obtained 
for a value of log(U)$\simeq\,-3$, in both regions. 
The ionization parameters for the NLR1 phases are log(U)=0.9$\rm{\pm}$0.2
and log(U)=0.3$\rm{\pm}$0.2 for NW and SE cones, respectively. These values are consistent with 
the parameters of the previous model.
The final fit can be seen in Figure \ref{fig:spectrumCloudy} and the 
best fit parameters and fluxes for each model component are given in Table \ref{tab:cloudyfit}.
 In order to show the confidence level of the fluxes, Fig. \ref{fig:isocontours} includes 
the iso-chi-squared flux contours of the normalizations of the two 
phases  (i.e. NLR1 and NLR2) of the reflected components for the NW (top) and SE (bottom) cones.
Similar result is obtained for the transmitted components.
The NLR1 phase mostly contributes to the X-ray spectrum at the region between 0.8-0.9 keV and 0.53-0.7 keV, 
whereas the NLR2 phase contributes at 0.52 and 0.7-0.85 keV.  
The lowest value of the ionization parameter is very similar to that required to fit the 
optical spectrum of Mrk~573 \citep{Kraemer09}. In that work, the authors claimed a three phase 
component to explain the optical emission line spectrum. The low-ionization gas accounts 
for the [OII]$\rm{\lambda\lambda}$3727\AA{} and [NII]$\rm{\lambda\lambda}$6548, 6584\AA{} emission, whereas the moderately ionized phase 
accounts for the [OIII]$\rm{\lambda\lambda}$5007\AA{}. Nonetheless, our NLR1 exhibits a value of U higher than 
their highly ionized phase. 
In addition, we expect a contribution of collisionaly ionized plasma (see Section \ref{sec:origin}) to
be present in the extended emission. However, given the fact that its 
contribution to the nuclear spectrum is about 6\%, we assumed its contribution to the
extended emission to be equal or smaller than in the nuclear case and did not try any
fit given the low count level in these regions.

In terms of flux, the NLR1 phase is the 64\% of the extended emission flux. 
Its contribution in the NW cone is higher than in the SE cone. Moreover, 
the reflection component dominates the NW cone whereas the transmission component dominates the SE one.
Tentatively, one could attribute this result to an orientation effect, being the NW cone located in the
farthest side and the opposite for the SE one. Indeed, 2-D spectroscopic observations \citep{Ferruit99}
indicates that at least part of the NW cone ionized gas is red-shifted with respect to the systemic velocity.
This could indicates that NW cone axis could be oriented behind the sky plane, although close to it. 
This orientation apparently contradicts the orientation proposed by
\citet{Tsvetanov92} based on reddening measurements, although we remark that 
their findings refer to the orientation with respect to the galaxy disk, not to the sky plane. 
Remarkably, the NLR2 phase accounts for the same flux in the NW and SE cones and 
nearly equal relative contribution of the reflection and transmission components. 
Thus, this low ionization phase seems to be uniformly distributed along the cone area.
 
\begin{figure}
\includegraphics[width=1.0\columnwidth]{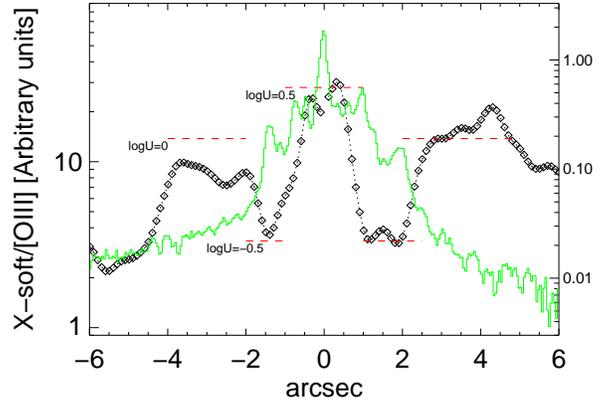}
\caption{\footnotesize{Variation of the ratio of soft X-ray to [O III] brightness profile in arbitrary units. 
The [O III] brightness profile variation is also overploted as a green histogram-like line. Positive x-axis values indicates the
NW orientation and negative values the SE one.
Horizontal dashed lines represent predictions from simple photoionization models at various values of the ionization
parameter.}
\label{fig:profX}}
\end{figure}

At this point, we may conclude that soft extended X-ray emission is then powered 
by features coming from ionized gas in two phases under NLR-like conditions.
Furthermore, one may question: is the [O III] extended emission structure powered
by the same mechanism? The similarity between the [O III] structures and the 
soft X-ray emission points to a common origin for both components. 
We have investigated the radial variation of the ratio between the soft X-ray and the [O III] line
emission. 
We represented in Figure \ref{fig:profX} the variation of the brightness ratio along the axis of the cone 
($\mathrm{PA}=122^\circ$). 
The brightness profiles were extracted using the IRAF {\it pvector} task. Before obtaining the ratio
we convolved the [O III] profile to obtain the resolution of 
X-ray data. As can be seen from Figure \ref{fig:profX} the
ratio soft-X/[O III] presents a non-uniform variation showing a maximum at the nucleus; it then  drops dramatically at the 
position of the [O III] arcs and returns at roughly half of the nuclear value
outside the [O III] arcs.  
A similar behaviour, namely a small variation of the ratio, has been reported by \citet{Bianchi06} for the case of NGC 3393. 
We have also compared  the radial variation with predictions from photoionization models in
Figure \ref{fig:profX}. We have used simple models assuming single plane--parallel slabs, constant density and radiation
bounded clouds.
The soft X-ray emission has been taken as the sum of the predicted values for the most intense
features identified in our X-ray spectra.
We have scaled the model predictions to the observed
nuclear values from the model with $\rm{log U = 0.5}$, which is close to the best fitting value derived above.
The predictions for different values of the ionization parameters are 
represented in Figure \ref{fig:profX}. It can be seen that a variation of $U_t$ by one order of magnitude is needed to 
reproduce the observed variation in the brightness ratio from the nucleus to the arcs, which could be attributed 
to a combination
of radiation dilution plus density enhancements at the arc positions.
Note, that these variations are qualitatively in agreement with our Cloudy model 
simulations in which two different U values are needed for SE and NW cones.
 A baseline model where the density decreases as
$r^{-2}$, as proposed by \citet{Bianchi06}, is compatible with our results, 
except for the regions close to [OIII] arcs.
It is very likely that simple photoionization 
models  are not adequate to reproduce the observed ionization 
variations, although a more sophisticated treatment must wait until 
higher quality X-ray spectroscopic observations become available. 
The fact that ionization- and matter-bounded clouds are likely constituents of the NLR has not been 
explored to explain emission lines in the soft-X range.

\section{The Nuclear Spectral Energy Distribution}

\begin{figure}
\includegraphics[width=1.0\columnwidth]{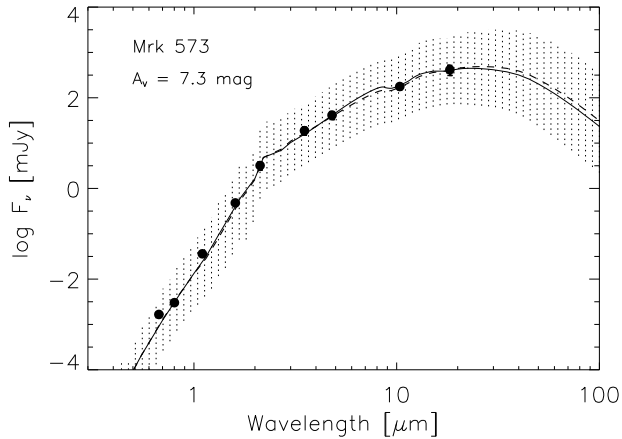}
\caption{High spatial resolution nuclear SED of Mrk 573 fitted with the clumpy torus models. Solid and dashed lines are the best fitting model 
and that computed with the median of each of the six parameters that describe the models (see \citealt{Ramos09b}). The shaded region indicates
the range of models compatible with a 68\% confidence interval around the median.\label{fig:opirSED}}
\end{figure}

Another important feature of the unified model is the optically thick torus. Under this scheme, Type-2 Seyferts 
like Mrk\,573 are obscured due to this material located along our line of sight. The best way to study and 
characterize the molecular torus is by modeling the nuclear spectral energy distribution (SED) of the sources. 
The near- and mid-IR nuclear emission of Seyfert galaxies is 
attributed to the reprocessing of the UV/X-ray nuclear radiation by
the toroidal dusty structure. For this reason, the infrared range is key to put constraints on torus modeling. However, 
in comparing the predictions of any torus model with observations, 
the small-scale torus emission must be isolated, in order
to avoid contamination from the host galaxy. For this reason, it is important 
to use high angular resolution data when trying to model the torus emission. 

We have tried here to explain the optical/infrared nuclear SED of Mrk 573 constructed with
high spatial resolution data (see Table \ref{tab:psf}) using an interpolated version of the recent models
for the clumpy torus scenario by \citet{Nenkova08a,Nenkova08b}. 
We have searched for the best fitting models using the Bayesian inference tool \textsc{BayesClumpy} \citep{Asensio09} 
The results of the fit are shown in Figure \ref{fig:opirSED}. 

Indeed, the nuclear SED of Mrk 573 has been previously fitted by \citet{Ramos09b} using this set of models and tools,
although the data points below 1~\micron\ were not included in their analysis. 
For using the optical photometry derived in this work, the clumpy 
torus model fit needs an additional extinction factor, which is 
included as a foreground extinction. The derived median value is $A_V = 7.3\pm0.5$ mag, which 
can be translated to a column density of $\rm{N_H = 1.39\times 10^{22}~\mathrm{cm}^{-2}}$. This value is
nicely consistent with that derived from the optical colour--colour maps (see Sect. \ref{sec:morpho}). 

The results of the fitting process are the probability distributions for the free 
parameters that describe the clumpy models (see \citealt{Ramos09b}). For Mrk 573, the median values of 
the parameters correspond to a torus width of 30$^\circ$, $N_0=4\pm1$ clouds in the equatorial direction, 
optical depth per single cloud $\tau_V \sim 60$, and $\mathrm{A_V^{LOS}}\sim 300$ mag (equivalent to 
$\mathrm{N_H^{LOS}}\sim 5.7\times10^{23}~\mathrm{cm}^{-2}$). These torus parameters are 
similar to those derived without including the optical data points by \citet{Ramos09b}. 

We have determined the torus luminosity integrating the corresponding emission from the torus model 
corresponding to the median value of the probability distribution 
of each parameter (dashed line in Figure \ref{fig:opirSED}).
The resulting value is L$_{bol}^{tor}=7.4 \times 10^{43}~\mathrm{erg s^{-1}}$.
The clumpy model fit yields the bolometric luminosity of the intrinsic
AGN, L$_{bol}^{AGN}=5.2 \times 10^{44}~\mathrm{erg s^{-1}}$ (the bolometric luminosities are good to a factor of 2). 
Combining this value with the torus luminosity, we derive the
reprocessing efficiency of the torus, which for Mrk 573 results to be quite low, about 14\%.

Moreover, we have derived the X-ray luminosity from the hard X-ray part, assuming a 
power law with photon index 1.8. The range from 2 to 6 keV was used to avoid, on the one hand the soft range,
which is attributed to emission lines, and the Fe K-line on the other. 
The galactic absorption plus an intrinsic absorption were included in the fitting process. The
intrinsic absorption results in a value of $\mathrm{N_H} = 7_{-5}^{+6} \times 10^{22}~\mathrm{cm}^{-2}$, which 
is compatible with the foreground extinction inferred from the colour maps and the clumpy torus modeling. 
Thus, the absorption corrected X-ray luminosity is 
$\mathrm{L_{2-10~keV}} =1.7 \times 10^{41}~\mathrm{erg\, s^{-1}}$.
This value is quite low when compared to the 
optical/infrared luminosity reprocessed by the blocking torus. However, it can be reconciled 
by taking into account that Mrk~573 has been classified as a \emph{Compton-thick} AGN \citep{Guainazzi05},
and its X-ray luminosity has to be corrected by a large factor to obtain the intrinsic luminosity, since the
column density is $\rm{N_H > 1.6\times 10^{24}~cm^{-2}}$ (such high value of the optical depth is also
consistent with the value derived above from the clumpy torus modeling). 
\citet{Panessa06} derived a factor of 60 comparing 
a small sample of {\it Compton-thick}~Type-2 Seyferts with a sample of Type-1 Seyferts, 
whereas \citet{Cappi06} derived a  factor about 100.
These values are in contrast with the
work of \citet{Gonzalez-Martin09}, who obtained a value of 42 using a sample of LINERs.  
In addition, we have to transform from X-ray to bolometric luminosity multiplying by a factor 
30 (\citealt{Risaliti04}; see also \citealt{Panessa06}). Thus, we obtain a value for the 
bolometric AGN luminosity in the range $\mathrm{L^{bol}_{AGN}=3.1-5.1 \times 10^{44}~erg s^{-1}}$,
which is in nice agreement with the value derived from the torus reprocessing, given the uncertainties involved. 
\citet{Kraemer09} derived a higher value for the bolometric luminosity ($3.2\times 10^{45}~erg s^{-1}$) based on
the [OIV]~25.9\micron\ luminosity, as measured by the Spitzer/IRS spectrum of Mrk~573 \citep{Melendez08a,Melendez08b}.
In any case Mrk~573 seems to be radiating near to the Eddington limit assuming the black hole mass
to be around $2\times 10^7 \mathrm{M_{\odot}}$ \citep{BianGu07}. This result adds support to the
re-classification of Mrk~573 as a hidden narrow-line Seyfert 1 \citep{Ramos08}.

\section{Conclusions and overall picture}

Mrk\,573 is a nearby optically classified Type-2 Seyfert, well-known for its extended circumnuclear emission-line regions. 
It is this extension and the proximity of the source that convert Mrk\,573 as one of the ideal cases to study 
this emission commonly 
found in Type-2 Seyfert galaxies. We combine RGS/\emph{XMM-Newton} and 
ACIS/\emph{Chandra} to achieve high spectral and spatial resolution in order to disentangle the emission 
mechanism of this extended emission. We also used optical and near-IR \emph{HST} data in order to compare with the X-ray 
data. The main results are:

\begin{itemize}
\item The soft X-ray emission is very complex, resembling that of the [O III] emission, as already 
reported. What constitutes a new result is that especially the NW structure is also very similar 
to that of [O III]/$\rm{H\alpha}$ emission. 
This suggests the same origin for the emission lines at optical  and the soft X-ray ranges.

\item Through X-ray spectroscopic analysis we have found that 
plasma excitation mechanism in the nuclear spectrum is mainly driven by photoionization from the 
central source, including a strong contribution from photoexcitation.  
A small contribution of collisionally ionized plasma is also needed to explain the emission 
line ratios shown by RGS spectra. This conclusion also agrees with the proposed
Cloudy simulations since the spectra could be interpreted as the combination of two different phases
of Cloudy models, with two different ionization parameters log(U)=1.23 and log(U)=0.13.

\item Based on the ACIS/\emph{Chandra} images 
and radial profiles along O VII triplet and O VIII RRC for cone-like structures 
we showed that O VIII RRC could be more relevant toward the NW cone, although 
the line ratios are formally compatible after including the error bars.

\item We have successfully modelled the cone-like emission using Cloudy simulations corresponding
to two phases of NLR conditions. The first phase shows different values of ionization parameter and different
contributions of the reflected and transmitted components for the NW (log(U)=0.9, reflected dominated) and SE
( log(U)=0.3, transmitted dominated) cones.
The second is an homogeneous phase with a lower ionization parameter (log(U)=-3) and
the same contribution of reflected and transmitted componets for NW and SE cones.

\item We have found a good agreement between the AGN bolometric luminosity derived from the 
hard X-ray luminosity and the one derived from the modeling of the optical/infrared  SED with
a clumpy torus model. 

\item From the extinction maps we have found that a dust lane crosses the nucleus in the N--S direction. 
This could be, in projection, perpendicular to the direction of the cone-like structure. The amount of 
extinction we have derived from this map is also consistent with that derived from the SED modeling. 
However, this must be related with the extended extinction since, in the inner parts, the AGN is hidden 
by a column density on the {\it Compton-thick} regime ($\rm{N_H>1.6\times 10^{24} cm-2}$).

\end{itemize}

 {\it Note added in manuscript:} After this paper was submitted to the journal a paper was published by 
\citet{Bianchi10} with data in common with our work. 
Most of their results are consistent with ours, although different approaches were used.
They fitted the \emph{XMM-Newton}/RGS spectra using Cloudy photoionization models. Their best fit 
was obtained by an hybrid model -photoionization + collisional excitation- where
the collisional phase contributes 1/3 of the flux in the band 0.5-0.8 keV, consistent with 
our work. 
Moreover, they claimed the need of two photoionization phases (log U = 0.3 and 1.8) to explain the ACIS/\emph{Chandra} 
spectrum 
in the range 0.4-7 keV, which is also in agreement with our results. 
The results on the extended emission cannot be compared since they do not distinguish between 
the two cone-like structures, extracting the spectrum from a circumnuclear annulus.


\acknowledgments
We thank to the anonymous referee for his/her helpful comments that have improved the 
final manuscript. 
Financial support by the grants AYA2006-09959,  AYA2007-60235 and AYA2008-06311-C02-01 from Plan Nacional de
Astronom{\ia}a y Astrof{\ia}sica is acknowledged. OGM acknowledges support by the EU FP7-REGPOT 206469 and ToK 39965 grants.
CRA acknowledges financial support from STFC PRDA (ST/G001758/1).
The authors acknowledge the Spanish Ministry of Science and Innovation (MICINN) through the Consolider-Ingenio 2010 Program 
grant CSD2006-00070: First Science with the GTC (http://www.iac.es/consolider-ingenio-gtc/).
OGM acknowledges  A. Nucita's help in using Cloudy models. 
The authors acknowledge Andr\' es Asensio Ramos for his valuable help and cooperation related 
with the use of BayesClumpy. 
Based on observations made 
with the NASA/ESA Hubble Space Telescope, and obtained from the 
Hubble Legacy Archive, which is a collaboration between the Space Telescope Science Institute (STScI/NASA), 
the Space Telescope European Coordinating Facility (ST-ECF/ESA) and the Canadian
Astronomy Data centre (CADC/NRC/CSA).






\end{document}